\documentclass[11pt, a4paper, oneside, reqno]{amsart}
\renewcommand\thesection{\arabic{section}}
\renewcommand\thesubsection{\thesection.\arabic{subsection}}

\makeatletter
\def\@settitle{\begin{center}%
		\baselineskip14\p@\relax
		\normalfont\LARGE\scshape\bfseries
		\@title
	\end{center}%
}
\makeatother

\makeatletter

\def\section{\@startsection{section}{1}%
	\z@{.7\linespacing\@plus\linespacing}{.5\linespacing}%
	{\normalfont\large\bfseries\centering}}

\def\subsection{\@startsection{subsection}{2}%
	\z@{.5\linespacing\@plus.7\linespacing}{.5\linespacing}%
	{\normalfont\large\bfseries}}

\def\subsubsection{\@startsection{subsubsection}{3}%
	\z@{.5\linespacing\@plus.7\linespacing}{.5\linespacing}%
	{\normalfont\itshape}}

\usepackage[colorlinks	= true,
			raiselinks	= true,
			linkcolor	= MidnightBlue,
			citecolor	= Mahogany,
			urlcolor	= ForestGreen,
			pdfauthor	= {Peyman Mohajerin Esfahani},
			pdftitle	= {},
			pdfkeywords	= {},   
			pdfsubject	= {},
			plainpages	= false]{hyperref}
\usepackage[usenames, dvipsnames]{color}
\usepackage{dsfont,amssymb,amsmath,subfig, graphicx,fancyhdr,mdframed}
\usepackage{amsfonts,dsfont,mathtools, mathrsfs,amsthm,wrapfig} 
\usepackage[]{siunitx}
\usepackage{ragged2e}
\let\labelindent\relax
\usepackage{enumitem}

\allowdisplaybreaks
\usepackage{algorithm}
\usepackage[noend]{algpseudocode}

\allowdisplaybreaks
\date{\today}

\patchcmd\@setauthors
  {\MakeUppercase{\authors}}
  {\authors}
  {}{}

\newtheorem{thm}{Theorem}[section]
\newtheorem{prop}{Proposition}

\newtheorem{lem}{Lemma}

\newtheorem{assum}{Assumption}
\newtheorem{prob}{Problem}

\newtheorem{defn}{Definition}

\theoremstyle{remark}
\newtheorem{rem}{Remark}

\theoremstyle{remark}

\theoremstyle{definition}

\newcommand{\EZ}[1][]{e}
\newcommand{\fEZ}[1][]{e}

\usepackage{changepage}

\begin{document}
	
		
		\title{ Robust Fault Estimators for Nonlinear Systems: \\
			    An Ultra-Local Model Design} 


		\author{Farhad Ghanipoor$^{1}$, Carlos Murguia$^{1}$, Peyman Mohajerin Esfahani$^{2}$, Nathan van de Wouw$^{1}$}
		
	\thanks{1. Department of Mechanical Engineering, Eindhoven University of Technology, The Netherlands}
	\thanks{2. Delft Center for Systems and Control, Delft University of Technology, The Netherlands}

		\maketitle
		
		\begin{abstract}                          
					This paper proposes a nonlinear estimator for the robust reconstruction of process and sensor faults for a class of uncertain nonlinear systems. The proposed fault estimation method augments the system dynamics with an ultra-local (in time) internal state-space representation (a finite chain of integrators) of the fault vector. Next, a nonlinear state observer is designed based on the known parts of the augmented dynamics. This nonlinear filter (observer) reconstructs the fault signal as well as the states of the augmented system. We provide sufficient conditions that guarantee stability of the estimation error dynamics: firstly, asymptotic stability (i.e., exact fault estimation) in the absence of perturbations induced by the fault model mismatch (mismatch between internal ultra-local model for the fault and the actual fault dynamics), uncertainty, external disturbances, and measurement noise and, secondly, Input-to-State Stability (ISS) of the estimation error dynamics is guaranteed in the presence of these perturbations. In addition, to support performance-based estimator design, we provide Linear Matrix Inequality (LMI) conditions for $\mathcal{L}_2$-gain and $\mathcal{L}_2-\mathcal{L}_\infty$ induced norm and cast the synthesis of the estimator gains as a semi-definite program where the effect of model mismatch and external disturbances on the fault estimation error is minimized in the sense of $\mathcal{L}_2$-gain, for an acceptable $\mathcal{L}_2-\mathcal{L}_\infty$ induced norm with respect to measurement noise. The latter result facilitates a design that explicitly addresses the performance trade-off between noise sensitivity and robustness against model mismatch and external disturbances. Finally, numerical results for a benchmark system illustrate the performance of the proposed methodologies. 
		\end{abstract}
	
	\color{black}
	\section{Introduction} \label{sec:intro}
		Process reliability is essential in many engineering systems, such as high-tech equipment, energy systems, automotive technology and health applications. Predictive maintenance technology is a key enabler for improving process reliability. A fundamental element for predictive maintenance is fault estimation. That is, we do not only need to know the presence and source of the fault (fault detection and isolation, see \cite{hwang2009survey,chen2012robust, ding2008model}, and references therein) but also its nature/severity (fault estimation). As an example, suppose the fault is small and/or slowly increasing in magnitude (slow compared to the system time scale). In this case, if accurate estimates of fault-induced signals are available the fault severity can be quantified and predictive maintenance can be scheduled accordingly. The latter is only possible if we estimate fault signals (at least their magnitude) using available information (inputs, measured outputs, and system models). Therefore, this paper focuses on the problem of fault estimation for a class of uncertain nonlinear systems.
	
	\emph{Existing Literature:} Available methods for fault estimation can be divided into three categories: 
	
	\emph{1)} Linear Systems: Numerous fault estimation methods have been developed for linear dynamical systems (see, e.g., \cite{liu2013sensor,liu2012fuzzy} for results on linear stochastic and switching systems). However, most practical systems, such as those in robotics, transportation systems, power networks, manufacturing, and water distribution, are nonlinear in nature. 
	
	\emph{2)} Nonlinear Systems/Linear Filters: Methods for nonlinear systems are still under development, see, e.g., \cite{zhu2015fault, ghanipoor2023linear, ossmann2016enhanced, de2001geometric}. For fault estimation in nonlinear systems, linear or nonlinear filters can be developed. Some of the existing literature on fault detection for nonlinear systems can be adapted to address the fault estimation problem \cite{varga2017solving}. For instance, \cite{esfahani2015tractable} provides a linear filter for fault detection of nonlinear systems in which the linear filter is designed by minimizing the nonlinearity effect on the filter output subject to a bound on the effects of fault on the filter output. In this result, by constraining the mapping from the fault to the filter output, the fault signal can be estimated. Similar results can be found in \cite{pan2021dynamic} where instead of minimizing the nonlinearity effect, the output mismatch of the actual and simulation-based system is minimized to provide robustifcation against model mismatch. 
	
	\emph{3)} Nonlinear Systems/Nonlinear Filters: For fault estimation in nonlinear systems, also nonlinear filters can be used \cite{Adaptive, veluvolu2011nonlinear, han2019intermediate, guzman2021actuator}. Most of the existing results construct nonlinear observers by incorporating nonlinear dynamics of the system as nonlinear filters for fault estimation. Because these results consider nonlinear dynamics in the filter structure, they can capture the behavior of nonlinear systems accurately and as a consequence provide better fault estimation. However, to provide such filters, due to nonlinearities, some assumptions on the class of systems and faults are required. Below we discuss some of these results. 
	
	As discussed in \cite{Adaptive}, the authors address the fault estimation problem for nonlinear systems with uniformly Lipschitz nonlinearities, process faults only (i.e., no sensor faults), and assume the so-called matching condition (the rank of the fault distribution matrix is invariant under left multiplication by the output matrix) is satisfied. An adaptive filter is provided that approximately reconstructs the actuator fault vector in this configuration. Although the matching condition makes the problem tractable, it significantly reduces the class of systems that can benefit from the results. In \cite{Phan2021}, a fault estimation scheme is introduced for both sensor and process faults using Nonlinear Unknown Input Observers (NUIO), adaptive Radial Basis Function Neural Networks (RBFNN), and assuming the matching condition is satisfied. The authors prove that their scheme provides boundedness of fault estimation errors.
	
	In \cite{zhu2015fault}, the matching condition does not need to be satisfied. However, they do not consider model uncertainty, external disturbances, and measurement noise. The authors consider Lipschitz nonlinearities, simultaneous sensor and process faults, and adopt a standard fault observability condition \cite{Patton_Input_Observability} on the linear part of the dynamics. Therein, the problem is tackled using the notion of intermediate observers, consisting of two dedicated observers, one that estimates the fault and the other the state. Their scheme guarantees bounded fault estimation errors. In \cite{van2022multiple}, simultaneous additive and multiplicative process faults are considered in the scope of discrete-time system models. They address the fault estimation problem by decoupling process nonlinearities and perturbations from the estimation filter dynamics and using regression techniques to approximately estimate fault signals. Decoupling nonlinearities leads to linear filters for which linear methods can be used to reconstruct fault signals. However, decoupling conditions impose strong assumptions on the system dynamics, which significantly limits the applicability of these results.
	
	 We remark that all the above-mentioned results for uncertain nonlinear systems guarantee \textit{approximate} reconstruction of fault vector only, i.e., they ensure bounded estimation errors, which, if small enough, still lead to a potentially good estimate of the true fault. Not having internal state-space representations of fault vector makes it challenging to enforce zero error fault estimation. We propose a fault estimation scheme for process and sensor faults that allows to guarantee zero error in the absence of uncertainty, external disturbances, and measurement noise for some classes of faults and ensures robust fault estimate in the presence of perturbations, all without \textit{requiring a matching condition}. 
	
	This scheme incorporates an internal representation of the fault vector, where we use the notion of \emph{ultra-local models} \cite{Flies_Ultra_Local, sira2018active} phenomenological models valid for short time intervals. We then extend the system dynamics with the (internal) ultra-local state-space of the fault vector to construct an augmented dynamics. Based on the known parts of the augmented dynamics, a nonlinear observer is proposed to estimate the states of the original system and the ultra-local fault (internal) system. 
	We derive the error dynamics of the observer in which the fault model mismatch (mismatch between actual internal system and its model), uncertainty mismatch (mismatch between actual uncertainty and its model), external disturbances, and measurement noise enter as external perturbations. The fault estimation problem is reformulated as a robust (against the mentioned perturbations) state estimation problem in the error dynamics.
	The main contributions of this paper are as follows:

	\begin{enumerate} [label=(\alph*)]
		\item 	 \textbf{\emph{Comprehensive Problem Setting:}}  Existing research on fault estimation for Lipschitz nonlinear systems has often skipped the comprehensive problem setting including time-varying process and sensor faults, modeling uncertainties, disturbances, and measurement noise \cite{han2019intermediate, Adaptive,van2022multiple}. Our contribution lies in presenting a fault estimation approach tailored to address all these challenges for Lipschitz nonlinear systems. 
		\item \textbf{\emph{Exact Fault Estimation Guarantee:}} 
		Existing studies for Lipschitz nonlinear systems primarily focus on achieving an \textit{approximate} reconstruction of the fault vector \cite{Adaptive, Phan2021,zhu2015fault}. In contrast, our proposed method stands out by offering a fault estimation framework that not only guarantees exact fault estimation for a class of time-varying faults (polynomial in time) when uncertainty, external disturbances, and measurement noise are absent. In addition, it ensures robust fault estimation with explicit, computable performance bounds in the presence of perturbations. 	
		\item \textbf{\emph{Optimization-Based Fault Estimation Scheme:}} Our approach introduces a computationally tractable algorithm to synthesize the fault estimator's design parameters. This is achieved through the solution of semi-definite programs, where we minimize the $\mathcal{L}_2$-gains from perturbations induced by fault, uncertainty model mismatches, and external disturbances to the fault estimation error. Additionally, we uphold stability conditions while simultaneously constraining desired upper bounds on the $\mathcal{L}_2-\mathcal{L}_\infty$ induced norms from measurement noise perturbations to the fault estimation error (Theorem 1). These computationally tractable design conditions provide a means to perform performance trade-off analyses in terms of robustness with respect to different disturbances/perturbations.
	\end{enumerate}

	This paper is a generalized version of the preliminary result published in the conference paper \cite{ghanipoor2022ultra}. Compared to \cite{ghanipoor2022ultra}, here we address a more general problem setting by considering state-dependent faults, external disturbances, model uncertainties and measurement noise. Furthermore, the fault estimate is robustified against perturbations induced by fault and uncertainty model mismatches, external disturbances, and measurement noise in the $\mathcal{L}_2$ sense for a desired worst-case noise amplification in the $\mathcal{L}_2-\mathcal{L}_\infty$ sense. 
	
	
	The remainder of this paper is organized as follows. In Section \ref{sec:problem formulation}, problem formulation is presented, first in a high level sense and then with precise mathematical details. The proposed method for the fault estimator design is described in Section \ref{sec:sol}. In Section \ref{sec:sim_results}, the method is applied to a benchmark example in simulation. Section \ref{sec:conclusion} presents the conclusion and final remarks.
	
	\textbf{Notation:}
	The symbol $\mathbb{R}^{+}$ denotes the set of nonnegative real numbers. The $n \times n$ identity matrix is denoted by $I_n$ or simply by $I$ if $n$ is clear from the context. Similarly, $n \times m$ matrices composed of only zeros are denoted by $0_{n \times m}$ or simply by $0$ when their dimensions are clear. For positive definite (semi-definite) matrices, we use the notation $P \succ 0$ $(P \succeq 0)$. For negative definite (semi-definite) matrices, we use the notation $P \prec 0$ $(P \preceq 0)$.
	The $\ell_2$ vector norm (Euclidean norm) and the matrix norm induced by the $\ell_2$ vector norm are both denoted as $||\cdot||$ and the $\ell_\infty$ vector norm is showed by $||\cdot||_\infty$. We use $\mathcal{L}_2(0,T)$ (or simply $\mathcal{L}_2$) to denote vector-valued functions $z:[0,T] \to \mathbb{R}^{k}$ satisfying $\|z(t)\|_{\mathcal{L}_2}^2 := \int_{0}^{T}\|z(t)\|^{2} d t<\infty$. For a vector-valued signal $f$ defined for all $t \geq 0$, $||f||_{\mathcal{L}_\infty} := \sup_{t \geq 0} ||f(t)||$ and $f^{(r)}$ shows the entry-wise $r^{th}$-time total derivative. For a differentiable function $W: \mathbb{R}^{n}  \to \mathbb{R}$ we denote by $\frac{\partial W}{\partial e}$ the row-vector of partial derivatives and by $\dot{W}(e)$ the total derivative of $W(e)$ with respect to time (i.e., $\frac{\partial W}{\partial e} \frac{de}{dt}$). We often omit time dependencies for notation simplicity. The notation $(f,d)$ stands for the column vector composed of the (vector or scalar) elements $f$ and $d$.

\section{Problem Formulation} \label{sec:problem formulation}
Consider the nonlinear system
	\begin{subequations} 	\label{eq:sys_all}
\begin{equation} \label{eq:sys}
	\left\{\begin{aligned}
		\dot{x} &=A x+ B_u u+ S_g g(V_g x, u,t) + S_\eta \eta(V_\eta x, u,t) +B_f f(x,u,t) \color{black}+ B_\omega \omega,  \\
		y&= C x + D_f f(x,u,t) \color{black}+ D_\nu \nu,
	\end{aligned}\right. 
\end{equation}
where $t \in \mathbb{R}^{+}$, $x \in {\mathbb{R}^{n}}$, $y \in {\mathbb{R}^{{m}}}$, and $u \in {\mathbb{R}^{{l}}}$ are time, state, measured output and known input vectors, respectively, and function $g: \mathbb{R}^{n_{v_g}} \times \mathbb{R}^{l} \times \mathbb{R}^{+} \to \mathbb{R}^{n_g}$ is a nonlinear known vector field. Function $\eta: \mathbb{R}^{n_{v_\eta}} \times \mathbb{R}^{l} \times \mathbb{R}^{+} \to \mathbb{R}^{n_\eta}$ denotes unknown modeling uncertainty. Signals $\omega: \mathbb{R}^{+}  \to \mathbb{R}^{n_\omega}$ and $\nu: \mathbb{R}^{+}  \to \mathbb{R}^{m_\nu}$ are unknown bounded disturbances; the former with unknown frequency range and the latter with high frequency content (e.g., related to measurement noise). Function $f: \mathbb{R}^{n} \times \mathbb{R}^{l} \times \mathbb{R}^{+}  \to \mathbb{R}^{n_f}$ denotes the unknown fault vector, which contains both process and sensor faults. Note that $f$ can represent any types of additive or multiplicative faults. Matrices $(A, B_u, S_g, V_g, S_\eta, V_\eta, B_f, B_\omega, C, D_f, D_\nu)$ are of appropriate dimensions, ${n},{m}, l, n_{v_g}, n_g, n_{v_\eta}, n_{\eta}, n_{\omega}, n_{f}, m_\nu \in \mathbb{N}$. Matrix $B_f$ denotes the process fault distribution matrix while matrix $D_f$ represents the contribution of the fault signal to sensor measurements. Matrices $S_g$ and $S_\eta$ are used to indicate in which equation(s) the nonlinearity $g$ and the uncertainty $\eta$ appear explicitly, and matrices $V_g$ and $V_\eta$ to indicate which states play a role in the nonlinearity and uncertainty, respectively.

The objective of this paper is to estimate the fault vector $f$ using the real time input, output data and the available known models. In the system \eqref{eq:sys}, clearly the uncertainty $\eta(\cdot)$ is unknown, which challenges fault estimation. We can either robustify the fault estimate error against the complete uncertainty or use an approximated model of it (obtained, e.g., using data-based methods) and robustify the fault estimate error agaisnt the remaining (smaller) uncertainty model mismatch. Without loss of generality, we assume that we can write $\eta(\cdot)$ in \eqref{eq:sys} as:
 \begin{equation} \label{eq:eta_lx}
	\begin{aligned}
		\eta(V_\eta x, u,t) = \eta_{l_x}(V_\eta x, u,t) + \delta \eta_{l_x}(V_\eta x, u,t), 
	\end{aligned}
\end{equation}
\end{subequations} 	
where $\delta \eta_{l_x}(V_\eta x, u,t) := \eta(V_\eta x, u,t) -  \eta_{l_x}(V_\eta x, u,t)$ and $\eta_{l_x}(\cdot)$ denotes any prior (possibly inaccurate) approximation, we may have of $\eta$. If we do not have such an approximated model, we just take $\eta_{l_x}=0$ an carry out further designs considering the complete $\eta$.

	{Let us state the assumptions on the system \eqref{eq:sys_all}, which stand throughout this paper.}

{\textbf{Regularity Assumptions:}
	The following assumptions are required to ensure that the problem is well-posed as is common in the existing literature \cite{sontag2008input,Keliris2017,han2019intermediate,Adaptive}:}
\vspace{-\baselineskip}
\begin{adjustwidth}{2em}{0pt}
	\begin{assum}\emph{\textbf{(State and Input Boundedness)}}  The state variable $x$ and the input $u$ are bounded over any finite time interval. 
		\label{assum:state_boundedness}
	\end{assum}
	
		\begin{assum}\emph{\textbf{($\mathcal{C}^r$ Fault Vector)}} The fault vector $f(x(t), {u}(t), t)$ in \eqref{eq:sys_all} is $r$ times differentiable with respect to time, i.e., the total time derivatives ${f}^{(1)}(x(t), {u}(t), t)$, ${f}^{(2)}(x(t), {u}(t), t)$, ..., $f^{(r)}(x(t), {u}(t), t)$ exist and are continuous. 
		\label{assum:fault_diff}
     	\end{assum}
	
	\begin{assum}\emph{\textbf{(Bounded Disturbances)}} The disturbance vectors $\omega$ and $\nu$ in \eqref{eq:sys_all} are bounded on any finite time interval, and $\nu$ is differentiable, i.e., the derivative $\dot{\nu}(t)$ with respect to time exists, is continuous, and is bounded over any finite time interval.
		\label{assum:dist_diff}
	\end{assum}
\end{adjustwidth}

We aim to robustify the fault estimate error against unknown bounded external disturbances (low or high frequency) and (fault and uncertainty) model mismatches. For fault estimation, we consider nonlinear filters with the following structure: 
\begin{equation} 		\label{eq:filter}
\left\{\begin{aligned} 
		\dot{z} =&h(z,u,y;\theta), \\
		\hat{f} =&\phi(z,y;\theta),
	\end{aligned}\right.
\end{equation}
where $z \in {\mathbb{R}^{n_z}}$ is the internal state of the filter with ${n_z} \in \mathbb{N}$. Functions $h: \mathbb{R}^{n_z} \times \mathbb{R}^{l} \times \mathbb{R}^{m} \to \mathbb{R}^{n_z}$ and $\phi: \mathbb{R}^{n_z} \times \mathbb{R}^{m} \to \mathbb{R}^{n_f}$ characterize the filer structure, $\theta$ denotes design parameters.

Define the fault estimate error as 
\begin{equation*} 
	e_f :=\hat{f}- f.
\end{equation*}
{Later, it will be shown that, for the fault estimator design proposed in Section \ref{sec:obsv}, the fault estimation error dynamics, explicated in Equation \eqref{eq:error_dynamics_combined} in Section \ref{sec:error_dynamics}, exhibits $(\delta \eta_{l_x}, \omega,  f^{(r)}, \nu, \dot{\nu})$ as a perturbation input. Now, having that in mind, we can state the problem we aim to solve at a high abstraction level.}
	\begin{prob}\emph{\textbf{(Fault Reconstruction - Abstract Level)}} Consider the uncertain nonlinear system \eqref{eq:sys_all} with known input and output signals, $u(t)$ and $y(t)$, and the nonlinear fault estimator filter \eqref{eq:filter}. For given $r$, design the filter parameters $\theta$ such that: \\
	\textbf{\emph{1) Stability:}} {The estimation error dynamics is input-to-state stable with respect to the perturbation input $(\delta \eta_{l_x}, \omega,  f^{(r)}, \nu, \dot{\nu})$\emph{;}}\\
	\emph{\textbf{2) Disturbance Attenuation:}} For $ \nu = 0$, the $\mathcal{L}_2$-gain from $(\delta \eta_{l_x}, \omega,  f^{(r)})$ to $e_f$ is bounded by some known $c_1 > 0$, for $t \geq 0$\emph{;}\\
	\emph{\textbf{3) Noise Rejection:}} For $ (\delta \eta_{l_x}, \omega,  f^{(r)})= 0$, the $\mathcal{L}_2-\mathcal{L}_\infty$ induced gain from $(\nu, \dot{\nu})$ to $e_f$ is bounded by some known $c_2 > 0$, for $t \geq 0$\emph{.}
	\label{prob:fault_estimation_high_level}
\end{prob}

We can further state the above problem as optimal filter design in the sense of disturbance attenuation or noise rejection. Now, we want to restate Problem \ref{prob:fault_estimation_high_level}, mathematically precisely. To this end, in what follows first, we discuss the fault estimator filter architecture. 

\subsection{Ultra Local Fault Representation}
	Let us introduce some preliminaries which are required to present the fault estimator filter architecture. {The fault $f(x(t), {u}(t), t)$ in (1a) is an implicit function of time, for all $x(t)$ and $u(t)$. For instance, the fault $f=  u(t)x(t)^2$, even though this $f$ is an explicit function of $(x(t),u(t))$, it can be considered to be an implicit function of time, given the fact that $x(t)$ and $u(t)$ are functions of time. Given this observation,} we can write an entry-wise $r$-th order Taylor time-polynomial approximation at time $t$ of $f$ as $\bar f = a_0 + a_1 t+ \dots + a_{r-1} t^{r-1}$ with coefficients $a_i \in {\mathbb{R}^{n_f}}, i= 0, \dots, r-1$. This model can be written in state-space:
\begin{equation} \label{eq:fault_model}
	\left\{\begin{aligned}
		\dot{\bar \zeta}_j &= \bar \zeta_{j+1},  \qquad 0 < j < r, \\
		\dot{\bar \zeta}_{r} &= 0,\\
		\bar f &= \bar \zeta_1,
	\end{aligned}\right.
\end{equation}
where $\bar \zeta_j \in {\mathbb{R}^{n_f}}$. Clearly, in the above model we have $\bar f^{(r)} = 0$, which might not be the case for actual fault signal $f$. Under Assumption \ref{assum:fault_diff}, the actual internal state-space representation of the fault $f$ is as follows:
  \begin{equation} \label{eq:fault_system}
  	\left\{\begin{aligned}
  		\dot{\zeta}_j &= \zeta_{j+1},  \qquad 0 < j < r, \\
  		\dot{\zeta}_{r} &= f^{(r)},\\
  		f &= \zeta_1,
  	\end{aligned}\right.
  \end{equation}
where $\zeta_j \in {\mathbb{R}^{n_f}}$. As you see, the accuracy of the approximated model \eqref{eq:fault_model} increases as $f^{(r)}$ goes to zero (entry-wise), and it is exact for $f^{(r)} = 0$ (since we have $\dot{\zeta}_r=\dot{\zeta}_1^{(r)}= {f}^{(r)}=0$, see \eqref{eq:fault_system}). Model \eqref{eq:fault_model} is used to construct a fault estimator filter that ultra-locally \cite{Flies_Ultra_Local}, \cite{sira2018active} acts as a self-updating polynomial spline approximating the actual value of the fault. To design such a filter, in the following section, we extend the system state, $x(t)$, with the states of the actual fault internal state $\zeta_j(t), j \in\left\{1, \ldots, r\right\}$, and augment the system dynamics in \eqref{eq:sys_all} with \eqref{eq:fault_system}. We then design a nonlinear filter (observer) for the augmented system to simultaneously estimate $x$ and $\zeta_j$ using model \eqref{eq:fault_model}. We remark that the number of the faults derivatives, $r$, added to the approximated model \eqref{eq:fault_model} (and \eqref{eq:fault_system}) is problem-dependent, and an optimal selection of $r$ would depend on the frequency characteristics of the fault. Increasing $r$ results in higher-dimensional augmented dynamics and thus high-dimensional observers as well. However, having larger observers also provides more degrees of freedom for optimal synthesis.
	
	\subsection{Augmented Dynamics}
	Based on the fault model introduced above, define the augmented state vector
	$x_a:= (x, \zeta_1,\zeta_2,\ldots,  \zeta_{r})$ 
	and write the augmented dynamics using \eqref{eq:sys_all} and \eqref{eq:fault_system} as
	\begin{subequations} 	\label{eq:augmented_fault}
		\begin{equation} 			\label{eq:augmented_system_fault}
			\begin{aligned}
				\left\{\begin{aligned}
					\dot{x}_{a} =&A_{a} x_{a}+ B_{u_a} u_a+ S_{g_a} g_a (V_{g_a} x_a, u_a, t) +B_{\omega_a} \omega_a, \\
					y = &C_{a} x_{a}+ D_{\nu} \nu,
				\end{aligned}\right.
			\end{aligned}
		\end{equation}
		where
		\begin{equation} 			\label{eq:augmented_matrices_fault}
			\begin{aligned}
				A_{a} &:=\left[\begin{array}{ccccc}
					A & B_f & 0 & \ldots & 0 \\
					0 & 0 & I_{n_f} & \ldots & 0 \\
					\vdots & \vdots & \vdots & \ddots & \vdots \\
					0 & 0 & 0 & \ldots & I_{n_f} \\
					0 & 0 & 0 & \ldots & 0
				\end{array}\right],  \quad
				 B_{u_a} :=	\left[\begin{array}{cc}
					{B}_u \\
					0 \\
				\end{array}\right], \quad
				S_{g_a} := \left[\begin{array}{cc}
					{S}_g & {S}_\eta\\
					0 & 0
				\end{array}\right], \\
				V_{g_a} &:= {\left[\begin{array}{cc} V_g  & 0 \\ V_\eta & 0 \end{array}\right]},   \quad
				{g}_a(\cdot) := {\left[\begin{array}{cc} g(\cdot)  \\ \eta_{l_x}(\cdot) \end{array}\right]}, \quad
				B_{\omega_a} :=	\left[\begin{array}{ccc}
					{S}_\eta & B_\omega & 0\\
					0 & 0  & 0\\
					0 & 0  & I_{n_f}\\
				\end{array}\right], \quad
		 \omega_a :=	\left[\begin{array}{c}
				\delta \eta_{l_x}(\cdot) \\
				\omega \\
				f^{(r)}(\cdot) \\
			\end{array}\right], \\
		 C_{a} &:= \left[\begin{array}{lll}	C & D_f	& 0 \end{array}\right], \qquad 	u_a := u.
			\end{aligned}
		\end{equation}
	\end{subequations}

Note that we have stacked the uncertainty model $\eta_{l_x}(\cdot)$ with the known nonlinearity $g(\cdot)$; however, if the uncertainty model is linear (see, e.g., the example in Section \ref{sec:sim_results}), it has to be stacked with the linear part of the augmented system $A_a$. Here we consider a more generic case where the uncertainty model is a globally Lipschitz nonlinear function. 

\begin{assum}\emph{\textbf{(Globally Lipschitz Nonlinearity)}} The function $g_a(V_{g_a} x_a, u_a, t)$ in \eqref{eq:augmented_system_fault} is globally Lipschitz uniformly in ${u}_a(t)$ and $t$, i.e., there exists a known positive constant $\alpha$ satisfying
	\begin{equation}
		\|g_a(V_{g_a} \hat{x}_a, u_a, t)-g_a(V_{g_a} x_a, u_a, t)\| \leq \alpha \|V_{g_a} (\hat{x}_a-x_a)\|,
		\label{eq:lipschitz} 
	\end{equation}
	for all $x_a,\hat{x}_a \in {\mathbb{R}^{n+rn_f}}$, ${u}_a \in {\mathbb{R}^{l}}$, and $t \in {\mathbb{R}^{+}}$.
	\label{assum:lipschitz}
\end{assum}

Note that in the above assumption, due to structure of $V_{g_a}$, $g_a(\cdot)$ is function $x$ (not $x_a$) and therefore the assumption is equivalent to $g(V_{g} x, u, t)$ and $\eta_{l_x}(V_{\eta} x, u, t)$ being globally Lipschitz uniformly in ${u}(t)$ and $t$ for all $x,\hat{x} \in {\mathbb{R}^{n}}$, ${u} \in {\mathbb{R}^{l}}$, and $t \in {\mathbb{R}^{+}}$.

	\subsection{Fault Estimator} \label{sec:obsv}
	In this section, considering the fault estimator general structure in \eqref{eq:filter}, inspired from observer-based approaches, we propose $h(\cdot)$ and $\phi(\cdot)$ as 
	\begin{subequations} \label{eq:observer}
		\begin{equation} 		\label{eq:nuio_sys_fault}
			\left\{\begin{aligned} 
				h(z,u,y;\theta) =& N z+G u_a+L y +M S_{g_a} {g}_a(V_{g_a}\hat{x}_a+ J\color{black}(y - C_{a} \hat{x}_a),u_a) \\
				\phi(z,y;\theta) =& \bar{C}(z-E \color{black} y) \\
			\end{aligned}\right.
		\end{equation}
		with $\hat{x}_a = z-E \color{black}y$,
		\begin{equation} 	\label{eq:c_bar_fault}
			\bar{C} := {\left[\begin{array}{lll}
					0_{n_f \times n} & I_{n_f} & 0_{n_f \times n_f(r-1)}
				\end{array}\right]},
		\end{equation}
		and matrices $\{N,G,L,M\}$ defined as
		\begin{equation} 		\label{eq:nuio_matrices_fault}
			\begin{aligned}
				N &:=M A_{a}- K \color{black} C_{a},
				&G&:=M  B_{u_a}, \\
				L &:=K \color{black}(I+C_{a} E \color{black})-M A_{a} E\color{black},
				&M &:=I+E \color{black}C_{a}.
			\end{aligned}
		\end{equation}
	\end{subequations}
	The algebraic relations in \eqref{eq:nuio_matrices_fault} are critical to nullify the effect of some signals on the fault estimation error. (see \eqref{eq:error_dynamics_split}). Furthermore, filter dimension $n_z$ equals to $n+rn_f$ and matrices $E$, $K$\color{black}, and $J$ \color{black} are filter gains to be designed which can be collected as $\theta = \{E,K,J\}$. In the following section, we analyze the fault estimator error dynamics.  
	
	\subsection{Fault Estimator Error Dynamics} \label{sec:error_dynamics}
	Consider the augmented state estimate $\hat{x}_a$ and define estimation error as
	\begin{equation} \label{eq:error_def}
		\begin{aligned}
		e:=&\hat{x}_a-x_a=z-x_a-E y=z-M x_a - E D_\nu \nu, \\
		e_f =& \hat{f} - f =  \bar{C}e.
		\end{aligned}
	\end{equation}
 The related estimation error dynamics can then be written as follows:
	\begin{equation} 	\label{eq:observer_dynamics_fault}
		\begin{aligned}
			\dot{e}&= N e+(N M+L C_{a}-M A_{a}) x_{a}+(G-M  B_{u_a}) u_a \\
			&+M S_{g_a}\big(g_a(V_{g_a}\hat{x}_{a} + J (y - C_{a} \hat{x}_{a}), u_a)-g_a(V_{g_a} {x}_{a}, u_a)\big) -M B_{\omega_a} \omega_a + (N E + L) D_\nu \nu - E D_\nu \dot{\nu}.
		\end{aligned}
	\end{equation}

	Given the algebraic relations in \eqref{eq:nuio_matrices_fault}, it can be verified that $G-M B_{u_a}= 0$, $N M+L C_{a}-M A_{a} = 0$, and $N E + L = K$. Therefore, \eqref{eq:observer_dynamics_fault} can be written as
		\begin{subequations}  \label{eq:error_dynamics_split}
		\begin{equation} 		\label{eq:error_dynamics_split_1}
			\begin{aligned}
				\dot{e} = &N e + M S_{g_a} \delta g_a -M B_{\omega_a} \omega_a +B_{\nu_a} \nu_a
			\end{aligned}
		\end{equation}
		with 
		\begin{equation} 		\label{eq:e_dynamics_martices}
			\begin{aligned} 
				\delta g_a &:= g_a(V_{g_a}\hat{x}_a+ J(y - C_{a} \hat{x}_a),{u})-g_a(V_{g_a} {x}_a, {u}), \\
					B_{\nu_a} &:= {\left[\begin{array}{cc}
						K D_\nu, & -E {D}_\nu
					\end{array}\right]}, 
				\qquad \nu_a := {\left[\begin{array}{cc}
						\nu^T, & \dot{\nu}^T
					\end{array}\right]^T}.
			\end{aligned}
		\end{equation}
	\end{subequations}

	By collecting all perturbations, error dynamics \eqref{eq:error_dynamics_split_1} can be written as
	\begin{subequations}  \label{eq:error_dynamics_combined}
		\begin{equation} \label{eq:error_dynamics_combined_1}
			\begin{aligned}
				\dot{e} &= N e + M S_{g_a} \delta g_a + \bar {B}_{\omega_a} \bar \omega_{a},
			\end{aligned}
		\end{equation}
		where 
		\begin{equation} \label{eq:omega_a_bar}
			\begin{aligned}
				\bar {B}_{\omega_a} &:= 
				{\left[\begin{array}{cc}
						-M B_{\omega_a}  & B_{\nu_a} 
					\end{array}\right]}, \qquad
				\bar \omega_a &:={\left[\begin{array}{cccc}
						\omega_a^T  &  \nu_a^T 
					\end{array}\right]^T} = {\left[\begin{array}{ccccc}
					\delta \eta_{l_x}^T  &  \omega^T  & f^{{(r)}^T}  &  \nu^T & \dot{\nu}^T
				\end{array}\right]^T}.
			\end{aligned}
		\end{equation}
	\end{subequations} 
	{
		Recall from Problem \ref{prob:fault_estimation_high_level} that we require the estimation error dynamics to have a certain Input-to State Stability (ISS) property. In the following definition, we define ISS for the estimation error dynamics \eqref{eq:error_dynamics_combined}. 
		\begin{defn} \emph{\textbf{(Input-to-State Stability)} The error dynamics \eqref{eq:error_dynamics_combined} is said to be ISS if there exist a class $\mathcal{K} \mathcal{L}$ function $\beta(\cdot)$ and a class $\mathcal{K}$ function $\mu(\cdot)$ such that for any initial estimation error $e(t_0)$ and any bounded input $\bar \omega_{a}$, the solution $e(t)$ of \eqref{eq:error_dynamics_combined} exists for all finite $t \geq t_{0}$ and satisfies
				\begin{equation}\label{ISS_def}
					\|e(t)\| \leq \beta\left(\left\|e\left(t_{0}\right)\right\|, t-t_{0}\right) + \mu( \sup _{t_0 \leq \tau \leq t}\|\bar \omega_a (\tau)\| ).  
			\end{equation}}
	\end{defn}}
	{Now, we have all the machinery required to restate Problem \ref{prob:fault_estimation_high_level} in a mathematically precise manner for the uncertain nonlinear system in \eqref{eq:sys_all}. }
	
	\begin{prob}\emph{\textbf{(Fault Reconstruction)}} Consider the uncertain nonlinear system \eqref{eq:sys_all} with known input and output signals, ${u}(t)$ and $y(t)$. Furthermore, consider the internal fault dynamics \eqref{eq:fault_system}, its Taylor approximation \eqref{eq:fault_model}, the augmented dynamics \eqref{eq:augmented_fault}, the nonlinear fault estimator filter \eqref{eq:filter} with functions defined in \eqref{eq:observer}, and let Assumption \ref{assum:lipschitz} be satisfied. Design the filter gain matrices $\theta = \{E,K,J\}$ such that we have: \\
		\textbf{\emph{1) Stability:}} {  
			The estimation error dynamics \eqref{eq:error_dynamics_combined} is ISS with respect to input $\bar \omega_a = (\omega_a, \nu_a)$\emph{;}}\\
		\emph{\textbf{2) Disturbance Attenuation:}} {for $\nu=\dot{\nu}=0$, it holds that
			\begin{equation} \label{eq:J1}
				J_1(\theta) := \sup_{(\delta \eta_{l_x}, \omega , f)} \frac{\|e_f\|_{\mathcal{L}_2}}{\| (\delta \eta_{l_x}, \omega , f^{(r)}) \|_{\mathcal{L}_2}}
			\end{equation}
			is bounded by some known $c_1 > 0$\emph{;}}\\[1 mm]
		\emph{\textbf{3) Noise Rejection:}} {for $\delta\eta_{l_x}=\omega=f^{(r)}=0$, it holds that
			\begin{equation} \label{eq:J2}
				J_2(\theta) := \sup_{\nu} \frac{\|e_f\|_{\mathcal{L}_2}}{\| (\nu, \dot{\nu}) \|_{\mathcal{L}_\infty}}
			\end{equation}
			is bounded by some known $c_2 > 0$\emph{.}}
		\label{prob:fault_estimation}
	\end{prob}

	Under Assumptions \ref{assum:state_boundedness}-\ref{assum:lipschitz}, Problem \ref{prob:fault_estimation} amounts to finding fault estimator gains that guarantee a bounded estimation error $e(t)$ in \eqref{eq:error_dynamics_combined}; for $\bar \omega_{a} = 0$, $e(t)$ goes to zero asymptotically (internal stability); when $\nu = 0$ the $\mathcal{L}_2$-gain of the mapping from $(\delta \eta_{l_x}, \omega , f^{(r)})$ to $e_f$ (the fault estimation error) is upper bounded by some $c_1>0$; and when $(\delta \eta_{l_x}, \omega , f^{(r)}) = 0$, the $\mathcal{L}_2-\mathcal{L}_\infty$ induced gain (or energy to peak norm \cite[pp. 78]{scherer2000linear}) from $(\nu, \dot{\nu})$ to $e_f$ is upper bounded by some $c_2>0$. 
	
	In what follows, we provide the solution to Problem \ref{prob:fault_estimation}.
	
	\section{Fault Estimator Design} \label{sec:sol}
	The solution to Problem \ref{prob:fault_estimation} is given in the following three subsections with the same order of the problem parts (i.e., 1. stability, 2. disturbance attenuation, and 3. noise rejection).
	
	\subsection{ISS Estimation Error Dynamics}
	In this section, we derive Linear Matrix Inequality (LMI) conditions for designing the matrices $\theta$ of the filter functions in \eqref{eq:observer}. As a stepping stone, we present a sufficient condition for asymptotic stability of the origin of the estimation error dynamics \eqref{eq:error_dynamics_combined} when the perturbation vector 
	$\bar \omega_{a} = (\omega_a, \nu_a)$
	 equals zero (internal stability). Moreover, we prove the boundedness of the estimation error in the presence of the perturbation using the input-to-state stability concept \cite{sontag2008input}.

We remark that $\bar \omega_{a}$ is a function of $x(t)$, $u(t)$ and $t$, so it is an implicit function of time. {Furthermore, note that $\bar \omega_a$ is bounded over any finite time interval due to Assumptions \ref{assum:state_boundedness} - \ref{assum:dist_diff} (see Remark \ref{rem:bounededness_assumptions} for further details). The boundedness of $\bar \omega_a$, together with the ISS of the estimation error dynamics in \eqref{eq:error_dynamics_combined}, implies boundedness of the estimation error over any finite time interval and the asymptotic stability of the origin of \eqref{eq:error_dynamics_combined} when  $\bar \omega_a = 0$. Moreover, if $\bar \omega_a$ is bounded uniformly in $t$; ISS guarantees the existence of an ultimate bound on the estimation error.}

\begin{rem}[\emph{\textbf{Boundedness Assumptions}}]
	{We have $\bar \omega_a = (\delta \eta_{l_x}, \omega, f^{(r)}, \nu, \dot{\nu})$. Note that, by the extreme value theorem \cite[Thm. 3.12]{protter2012first}, the uncertainty model mismatch $\delta \eta_{l_x}(x(t),u(t),t)$ and the fault model mismatch $f^{{(r)}}(x(t), u(t), t)$ are bounded over any finite time interval due to continuity and bounded driving terms (Assumptions \ref{assum:state_boundedness} and \ref{assum:fault_diff}). Moreover, the other terms in $\bar \omega_a$ (i.e., $\omega(t)$, $\nu(t)$, and $\dot{\nu}(t)$) are also bounded by Assumption \ref{assum:dist_diff}. Therefore, $\bar \omega_a$ is bounded over any bounded time interval under the required assumptions.
		\label{rem:bounededness_assumptions}}
\end{rem}

	The next proposition formalizes an LMI condition that guarantees an ISS estimation error dynamics \eqref{eq:error_dynamics_combined} with respect to input $\bar \omega_a$. 
	
	\begin{prop}\emph{\textbf{(ISS Estimation Error Dynamics)}} Consider the error dynamics \eqref{eq:error_dynamics_combined} {and let Assumption \ref{assum:lipschitz} holds with Lipschitz constant $\alpha$.} Suppose there exist matrices $ P \in {\mathbb{R}^{n_z \times n_z}},$ with $P \succ0, R \in {\mathbb{R}^{n_z \times m}}, Q \in {\mathbb{R}^{n_z \times m}}$,  and $J \in {\mathbb{R}^{n_{v_g} \times m}}$ satisfying the matrix inequality
		\begin{subequations} 
			\begin{equation} \label{eq:stability_lmi}
				\left[\begin{array}{cc}
					X_{11} & X_{12} \\
					* & -I
				\end{array}\right] \prec 0,
			\end{equation}
			where, matrices $X$ and $X_{12}$ defined as
			\begin{equation} \label{eq:X12}
				\begin{aligned}
					X_{11} :=& S_{11}+\alpha (V_{g_a}^{T} V_{g_a} -V_{g_a}^{T}  J  C_{a} - C_{a}^{T}  J^{T}  V_{g_a}) \\
					X_{12} :=&  {\left[\begin{array}{c}
							\sqrt{2\alpha}\big((P+R C_{a})S_{g_a}\big)^T \\  \sqrt{\alpha}(J C_{a})^T  
						\end{array}\right]^T}
									\end{aligned}
			\end{equation}
		with
		\begin{equation} \label{eq:s11}
			\begin{aligned}
				S_{11} := &A_{a}^{\top}  P +A_{a}^{\top} C_{a}^{\top} R^{\top}-C_{a}^{T}  Q^{T} + P A_{a} +R C_{a} A_{a} -  Q C_{a}, \\
				\end{aligned}
			\end{equation}
		$\alpha$ from \eqref{eq:lipschitz}, and the remaining matrices in \eqref{eq:augmented_matrices_fault}. Then, the estimation error dynamics in \eqref{eq:error_dynamics_combined} is ISS with respect to input $\bar \omega_a$. Moreover, when $\alpha = 0$ (no nonlinearity) the condition in \eqref{eq:stability_lmi} transforms to 
		\begin{equation} \label{eq:stability_lmi_linear}
				S_{11}  \prec 0,
		\end{equation}
			\end{subequations} 
		that is a necessary and sufficient condition for ISS with respect to input $\bar \omega_a$.
		\label{propos:stability}
	\end{prop}
	\emph{\textbf{Proof}}: The proof can be found in Appendix \ref{ap:propos1_proof}. 
	\hfill $\blacksquare$
	
	\begin{rem}[\emph{\textbf{LMI Feasibility}}]
		If ${D}_f$ in \eqref{eq:sys} is full row rank (i.e., there are as many sensor faults, $f_y$, as sensors), the LMI in \eqref{eq:stability_lmi} is always infeasible (observability is lost) \emph{\cite{edwards2006comparison, edwards2000sliding, frank1989robust}}. The standard practice to circumvent this issue is to assume $\text{rank}[{D_f}] < m$ \emph{\cite{edwards2006comparison, edwards2000sliding, frank1989robust}}. Note that this is only a necessary condition and it does not guarantee the LMI in \eqref{eq:stability_lmi} to be feasible (this has to be checked on a case-by-case basis).
		\label{rem:feasibility}
	\end{rem}

		In what follows, we shift our attention from the error dynamics \eqref{eq:error_dynamics_combined} to its equivalent  \eqref{eq:error_dynamics_split}, to characterize fault estimation robustness against disturbances with unknown frequency range $(\delta \eta_{l_x}, \omega, f^{(r)})$ (external disturbances, and fault and uncertainty model mismatches) and disturbances with high frequency content $(\nu, \dot{\nu})$ (measurement noise and its derivative). Because $(\delta \eta_{l_x}, \omega, f^{(r)})$ and $(\nu, \dot{\nu})$ have different frequency characteristics, robustness strategies for each of these perturbations are different. 
	
	\subsection{$\mathcal{L}_2$ Performance Criterion}
	To maximize the performance of the reconstruction scheme, we seek to minimize the effect of $(\delta \eta_{l_x}, \omega, f^{(r)})$ (treated as an arbitrary energy bounded external disturbance) on the fault estimation error $e_f$. We assume $\nu$ is zero (consequently $\nu_a = 0$) in the error dynamics \eqref{eq:error_dynamics_split} since we apriori know that the measurement noise has high frequency content and thus we consider the effect of $\nu_a$ in the next section. To this end, we seek to minimize the $\mathcal{L}_2$-gain from $(\delta \eta_{l_x}, \omega, f^{(r)})$ to the fault estimation error $e_f$. We could use the ISS formulation in Proposition \ref{propos:stability} to cast an optimization problem where we minimize the ISS gain and treat the LMI in \eqref{eq:stability_lmi} as an optimization constraint. By doing so, we would be reducing the effect of $(\delta \eta_{l_x}, \omega, f^{(r)})$ on the complete vector of estimation errors $e$ (state, and fault and fault derivatives estimation errors). Note, however, that the filter's purpose is to reconstruct fault vector only, so the performance in state estimation and the error of higher-order fault derivatives is not relevant. 
	\begin{defn} \textbf{\emph{($\mathcal{L}_2$-gain~\cite{van19922})}}
		We say that the estimation error dynamics \eqref{eq:error_dynamics_split}, assuming $\nu = 0$ with input $\omega_a(t) = (\delta \eta_{l_x}, \omega, f^{(r)})$ and output $e_f(t)$ (fault estimation error as in \eqref{eq:error_def}) has a $\mathcal{L}_2$-gain less than or equal to $\lambda$ if the following inequality is satisfied
		\begin{equation*}
			||e_f(t)||_{\mathcal{L}_2} := \int_{0}^{T}\|e_f(t)\|^{2} d t \leq \lambda^{2} \int_{0}^{T}\|\omega_a(t)\|^{2} d t,
		\end{equation*}
		for all $T \geq 0$ and $\omega_a(t) \in \mathcal{L}_2(0,T)$.
		\label{def:l2-gain}
	\end{defn}
	
	The following proposition formalizes an LMI-based condition guaranteeing that \eqref{eq:error_dynamics_split} has the finite $\mathcal{L}_2$-gain property from $(\delta \eta_{l_x}, \omega, f^{(r)})$ to the fault estimation error $e_f$ (see Definition \ref{def:l2-gain}).
	\begin{prop}[$\mathcal{L}_2$-gain LMI]
		{Consider the error dynamics \eqref{eq:error_dynamics_split} and let Assumption \ref{assum:lipschitz} holds with Lipschitz constant $\alpha$.} Suppose there exist matrices $P\succ0, R,$ $Q$, $J$, and scalar $\rho \geq 0$ satisfying
		\begin{subequations} 
			\begin{align}
				&\left[\begin{array}{ccc}
					X_{11}+ a \bar{C}^{T} \bar{C} & -(P + R C_a) B_{\omega_a} & X_{12} \\
					* & -{\rho} a I  & 0 \\
					* & * & -I
				\end{array}\right]\preceq0, \label{eq:l2_gain_lmi}
			\end{align}
		for some given $a>0$, $X_{11}, X_{12}$ as defined in \eqref{eq:X12}, $\bar{C}$ in \eqref{eq:c_bar_fault}, and the remaining matrices in \eqref{eq:augmented_matrices_fault}. Then, $J_1(\cdot)$ in \eqref{eq:J1} is upper bounded by $\sqrt{\rho}$, i.e., the $\mathcal{L}_2$-gain of \eqref{eq:error_dynamics_split} with $\nu = 0$ from $(\delta \eta_{l_x}, \omega, f^{(r)})$ to the fault estimation error $e_f$ is upper bounded by $\sqrt{\rho}$. Moreover, when $\alpha = 0$ (no nonlinearity) the condition in \eqref{eq:l2_gain_lmi} transforms to 
		\begin{equation} \label{eq:l2_gain_lmi_linear}
\begin{aligned}
	&\left[\begin{array}{cc}
		S_{11} + \bar{C}^{T} \bar{C} & (P + R C_a) B_{{d}_{1}} \\
		* & -{\rho}  I   \\
	\end{array}\right]\preceq0,\\
\end{aligned}
		\end{equation}
	\end{subequations} 
that is a necessary and sufficient condition for $\sqrt{\rho}$ to be the upper bound for the $\mathcal{L}_2$-gain from $(\delta \eta_{l_x}, \omega, f^{(r)})$ to the fault estimation error $e_f$.
\label{propos:L_2_gain}
	\end{prop}
	\emph{	\textbf{Proof}:}
	The proof can be found in Appendix \ref{ap: l_2_gain}.
	\hfill $\blacksquare$ 
	
		\subsection{$\mathcal{L}_2-\mathcal{L}_\infty$ Induced Norm Performance Criterion}
	The other terms affecting the fault reconstruction are measurement noise and its derivative with high frequency content $(\nu, \dot{\nu})$. Therefore, we assume disturbances with unknown frequency range $(\delta \eta_{l_x}, \omega, f^{(r)}) = 0$ in the error dynamics \eqref{eq:error_dynamics_split} and seek to minimize the effect of disturbances with high frequency content $(\nu, \dot{\nu})$ on the fault estimate error $e_f$. Here, we should select an appropriate performance criterion to characterize the effect of $(\nu, \dot{\nu})$ on the estimation error dynamics \eqref{eq:error_dynamics_split}. Because $(\nu, \dot{\nu})$ consists of disturbances with bounded energy, we consider $\mathcal{L}_2$-norm for that and since it leads to abrupt changes in fault estimates $e_f$, we use $\mathcal{L}_\infty$-norm for fault estimation error $e_f$ to have a filtering effect. In other words, we minimize the effect of abrupt changes and push the maximum amplitude in the fault estimate error signal down to have a smooth fault estimate. Therefore, we seek to minimize the $\mathcal{L}_2-\mathcal{L}_\infty$ induced norm from $(\nu, \dot{\nu})$ to fault estimation error $e_f$. Moreover, one can note that the $\mathcal{L}_2-\mathcal{L}_\infty$ induced norm is widely used in the literature for filtering (See \cite{shen2016extended,ahn2013l_,shen2015reliable,song2016l2,zhang2017h}). 
	
	\begin{defn} \textbf{($\mathcal{L}_2-\mathcal{L}_\infty$ Induced Norm)}
		We say that the estimation error dynamics \eqref{eq:error_dynamics_split} assuming $(\delta \eta_{l_x}, \omega, f^{(r)}) = 0$ with input $\nu_a(t) = (\nu, \dot{\nu})$ and output $e_f(t)$ (fault estimation error as in \eqref{eq:error_def}) has a $\mathcal{L}_2-\mathcal{L}_\infty$ induced norm less than or equal to $\gamma$ if the following inequality is satisfied
		\begin{equation*}
			||e_f||_{\mathcal{L}_\infty} := \sup_{t \geq 0} \|e_f(t)\|^{2} \leq \gamma^{2} \int_{0}^{\infty}\|\nu_a (t)\|^{2} d t,
		\end{equation*}
		for all $\nu_a (t) \in \mathcal{L}_2(0,\infty)$.
		\label{def:linf_l2}
	\end{defn}
	
	In the following proposition, we give a Lyapunov-based sufficient LMI condition for having a bounded $\mathcal{L}_2-\mathcal{L}_\infty$ induced norm of the mapping from $(\nu, \dot{\nu})$ to the fault estimation error $e_f$.
	
	\begin{prop}[\textbf{$\mathcal{L}_2-\mathcal{L}_\infty$ Induced Norm LMI}]
			{Consider the error dynamics \eqref{eq:error_dynamics_split} and let Assumption \ref{assum:lipschitz} holds with Lipschitz constant $\alpha$.} Suppose there exist matrices $P\succ0, R,$ $Q$, $J$, and scalar $\sigma \geq 0$ satisfying
		\begin{subequations} \label{eq:lmi_variables_prop3}
			\begin{align}
				&\left[\begin{array}{cccc}
					X_{11} & H_{12} & 0 & X_{12}\\
					* & -b^2 I  &  T_\nu^T J^T & 0\\
					* & * & -I & 0\\
					* & * & * & -I
				\end{array}\right]\preceq0,  \label{eq:l2_linf_1}\\
				&\left[\begin{array}{cc}
					P	 & \bar{C}^T \\
					* & \sigma I
				\end{array}\right] \succeq 0, \label{eq:l2_linf_2}\\
				&H_{12} :=[\begin{array}[]{ll}  Q D_\nu , & -R D_\nu \end{array}] \label{eq:h12}, \\ 
				&T_\nu = [\begin{array}[]{ll}  D_\nu & 0 \end{array}] \label{eq:T_nu}, 
			\end{align}
		for some given scalar $b$, $X_{11}, X_{12}$ as defined in \eqref{eq:X12}, $\bar{C}$ in \eqref{eq:c_bar_fault}, and the remaining matrices in \eqref{eq:augmented_matrices_fault}. Then, $J_2(\cdot)$ in \eqref{eq:J2} is upper bounded by $\sqrt{b \sigma}$, i.e.,  the $\mathcal{L}_2-\mathcal{L}_\infty$ induced norm of \eqref{eq:error_dynamics_split} with $ \omega_a = 0$ from $(\nu, \dot{\nu})$ to the fault estimation error $e_f$ is upper bounded by $\sqrt{b \sigma}$. Moreover, when $\alpha = 0$ (no nonlinearity) the condition in \eqref{eq:l2_linf_1} reduces to 
		\begin{equation} \label{eq:l2_linf_gain_lmi_linear}
			\begin{aligned}
				&\left[\begin{array}{cc}
					S_{11} & H_{12} \\
					* & - I   \\
				\end{array}\right]\preceq0,\\
			\end{aligned}
		\end{equation}
	\end{subequations} 
that is a necessary and sufficient condition for $\sqrt{\sigma}$ to be the upper bound for the $\mathcal{L}_2-\mathcal{L}_\infty$ induced gain from $(\nu, \dot{\nu})$ to the fault estimation error $e_f$.
		\label{propos:L2_Linf_norm}
	\end{prop}
	\emph{	\textbf{Proof}:}
	The proof can be found in Appendix \ref{ap:l2_linf_norm}.
	\hfill $\blacksquare$
	
		\begin{rem}[\emph{\textbf{Extension}}]
			Although due to the nature of signals unknown frequency range $(\delta \eta_{l_x}, \omega, f^{(r)})$ and high frequency content $(\nu, \dot{\nu})$, we have proposed $\mathcal{L}_2$-gain and $\mathcal{L}_2-\mathcal{L}_\infty$ induced gain results, respectively, we can apply $\mathcal{L}_2-\mathcal{L}_\infty$ induced norm norm to $(\delta \eta_{l_x}, \omega, f^{(r)})$, or the other way around. For example, applying $\mathcal{L}_2-\mathcal{L}_\infty$ induced norm to $(\delta \eta_{l_x}, \omega, f^{(r)})$ can be useful when an uncertainty with high-frequency content appears in $\delta \eta_{l_x}$. This case is, however, not considered here.
		\end{rem}
	
			\begin{rem}[\emph{\textbf{Exact Estimation}}]
			The developed methodology can guarantee zero estimation error for zero $\bar \omega_{a} = (\delta \eta_{l_x}, \omega, f^{(r)}, \nu, \dot{\nu})$, i.e., when the $r^{th}$-time derivative of the fault vector vanishes (i.e., time polynomial signals with a degree less than r), and the disturbances (the uncertainty model mismatch $\delta \eta_{l_x}$, the external disturbance $\omega$, and the measurement noise $\nu$) are zero. {This follows directly from the ISS property that the origin of the estimation error dynamics in \eqref{eq:error_dynamics_combined} is asymptotically stable if $\bar \omega_a = 0$.} See simulation results in \cite{ghanipoor2022ultra}.
		\end{rem} 
		
		\begin{rem}[\emph{\textbf{Fault Internal Model}}]
			It is worth highlighting that the fault internal model introduced in \eqref{eq:fault_model}, with $r$ as a design parameter, allowing the addition of as many terms as needed from the Taylor series, can be generalized for a wide class of fault signals. However, if there exists prior knowledge about the fault signal, such as a specific known frequency, the internal fault model can be adapted accordingly \cite{dong2023robust}. By doing so, the proposed method can guarantee exact fault estimation for the class of faults for which the model is exact.
			\label{rem:internal_model}
		\end{rem}
		
Propositions \ref{propos:stability}-\ref{propos:L2_Linf_norm} provide sufficient conditions that we exploit to solve Problem \ref{prob:fault_estimation} (filter synthesis) in what follows. So far, we have presented analysis tools to characterize the performance of a given filter of the form \eqref{eq:filter} with functions in \eqref{eq:observer}. In the following section, we provide a tool to design the filter matrices $\theta = \{E,K,J\}$ in \eqref{eq:observer}, in an optimal way in the sense of achieving a desired trade-off between the $\mathcal{L}_2$-gain and $\mathcal{L}_2-\mathcal{L}_\infty$ induced-gain introduced above. 

\subsection{Optimal Fault Estimator Design}
	Using Proposition \ref{propos:L_2_gain}, we can formulate a semi-definite program where we seek to minimize the $\mathcal{L}_2$-gain from unknown frequency range disturbances $\omega_a = (\delta \eta_{l_x}, \omega, f^{(r)})$ to fault estimation error $e_f$. Similarly, using Proposition \ref{propos:L2_Linf_norm}, we can formulate another semi-definite program where we seek to minimize the $\mathcal{L}_2-\mathcal{L}_\infty$ induced norm from disturbances with high frequency content $\nu_a = (\nu, \dot{\nu})$ to $e_f$. However, in the presence of both unknown perturbations ($\omega_a$ and $\nu_a$), the $\mathcal{L}_2$-gain and $\mathcal{L}_2-\mathcal{L}_\infty$ can not be minimized simultaneously due to conflicting objectives. To attenuate the effect of high-frequency disturbances on the estimation error, a relatively slow (low-gain) filter is required, which does not react to every small and fast change in the measured output. On the other hand, to reduce the effect of $\omega_a$ on the estimation error, a high-gain filter is preferred, which tries to estimate the fault as accurately as possible. It follows that there is a trade-off between estimation performance the noise sensitivity. 
	To address this trade-off, a convex program can be proposed where we seek to minimize the $\mathcal{L}_2$-gain and constrain the $\mathcal{L}_2-\mathcal{L}_\infty$ induced gain. The same tools allow minimizing the $\mathcal{L}_2-\mathcal{L}_\infty$ norm for a constrained $\mathcal{L}_2$-gain (as a dual problem). Moreover, we add the ISS LMI in \eqref{eq:stability_lmi} as a constraint to these programs to enforce that the resulting filter also guarantees boundedness for bounded perturbations and asymptotic stability for vanishing $\omega_a$ and $\nu_a$ (as having bounded signal norms does not guarantee having bounded filter trajectories \cite{khalil2002nonlinear}).

	\begin{thm}\textbf{\emph{(Optimal Fault Estimator)}}
		Consider the augmented dynamics \eqref{eq:augmented_fault}, the fault estimator filter \eqref{eq:filter} with $h(\cdot)$ and $\phi(\cdot)$ as defined in \eqref{eq:observer}, and the corresponding estimation error dynamics \eqref{eq:error_dynamics_split}-\eqref{eq:error_dynamics_combined}. Let Assumptions \ref{assum:state_boundedness}-\ref{assum:lipschitz} be satisfied. To design the parameters of the fault estimator, solve the following convex program
		\begin{equation}
			\begin{array}{cl}
				\min \limits_{P, R, Q, J, \rho, \sigma}  & \rho \\
				\text{s.t.} & 						\vspace{1 mm}
				\left[\begin{array}{cc}
					X_{11} & X_{12} \\
					* & -I
				\end{array}\right] \prec 0,\\
				& \left[\begin{array}{ccc}
					X_{11}+ a \bar{C}^{T} \bar{C} & -(P + R C_a) B_{\omega_a} & X_{12} \\
					* & -{\rho} a I  & 0 \\
					* & * & -I
				\end{array}\right]\preceq0, \\[5mm]
				&\left[\begin{array}{cccc}
					X_{11} & H_{12} & 0 & X_{12}\\
					* & -b^2 I  &  T_{\nu}^T J^T & 0\\[5mm]
					* & * & -I & 0\\
					* & * & * & -I
				\end{array}\right]\preceq0, \\[5mm]
				&\left[\begin{array}{cc}
					P	 & \bar{C}^T \\
					* & \sigma I
				\end{array}\right] \succeq 0,\\
				& P \succ 0, \quad \rho, \sigma \geq 0, \quad \sigma \leq \sigma_{x_{max}}
			\end{array}
			\label{eq:mimization}
		\end{equation}
		with given scalars $a, \sigma_{x_{max}} > 0$ and $b$, $X_{11}, X_{12}$ as defined in \eqref{eq:X12}, $B_{\omega_a}$ and $C_a$ in \eqref{eq:augmented_matrices_fault}, $H_{12}$ in \eqref{eq:h12}, $T_{\nu}$ in \eqref{eq:T_nu}, and $\bar{C}$ in \eqref{eq:c_bar_fault}. Denote the optimizers as $P^\star$, $R^\star$, $Q^\star$, $J^\star$, $\rho^\star$ and $\sigma^\star$. Then, the following parameters of \eqref{eq:observer}, $\theta= \theta^\star = \{E^\star = P^{\star^{-1}} R^\star ,K^\star = P^{\star^{-1}} Q^\star,J^\star\}$ guarantees the following:
		\begin{enumerate}
			\item The estimation error dynamics in \eqref{eq:error_dynamics_combined} is ISS with respect to input $\bar \omega_a = (\omega_a, \nu_a)$. In addition, the ISS property guarantees the asymptotic stability of the origin of the estimation error dynamics for $\bar \omega_a = 0$. Moreover, if $\bar \omega_a$ is bounded uniformly in $t$; ISS implies the existence of an ultimate bound on the estimation error;
			\item $J_1(\cdot)$ in \eqref{eq:J1} is upper bounded by $\sqrt{ \rho^\star}$, i.e., the $\mathcal{L}_2$-gain of \eqref{eq:error_dynamics_split} with $\nu = 0$ from $\omega_a = (\delta \eta_{l_x}, \omega, f^{(r)})$ to the fault estimation error $e_f$ is upper bounded by $\sqrt{ \rho^\star}$. 
			\item $J_2(\cdot)$ in \eqref{eq:J2} is upper bounded by $\sqrt{b \sigma^\star}$, i.e., the $\mathcal{L}_2-\mathcal{L}_\infty$ induced norm of \eqref{eq:error_dynamics_split} with $\omega_a= 0$ from $\nu_a = (\nu, \dot{\nu})$ to $e_f$ is upper bounded by $\sqrt{b \sigma^\star}$.
		\end{enumerate}
		\label{theorem:fault_estimation}
	\end{thm}
	\emph{\textbf{Proof}:} 
	Theorem \ref{theorem:fault_estimation} follows from the above discussion and Propositions \ref{propos:stability}-\ref{propos:L2_Linf_norm}.
	\hfill $\blacksquare$

Note that, for numerical tractability, we replace $X_{11}$ in the first strict inequality and $P \succ 0$ in the last strict inequality with $X_{11}+ \epsilon I$  (it becomes non-strict inequality) and $P- \epsilon I \succeq 0$, with a given $\epsilon > 0$, respectively (See \cite{ghanipoor2022ultra} for the relation of $\epsilon$ and ISS-gain). {Furthermore, the scalar parameters $a$ and $b$ in Theorem 1 are tuned for the minimal $\mathcal{L}_2$-gain with respect to disturbances $\omega_a$ for an acceptable $\mathcal{L}_2-\mathcal{L}_\infty$-gain for noise $\nu_a$ by a line search. To facilitate the implementation of the proposed scheme, the steps to implement it are given as follows:}
\begin{enumerate}
	\item Reformulate the system dynamics in the form of (1);
	\item Select the number of fault derivative augmentation $r$ in (4);
	\item  Construct the known part of augmented dynamics (5) based on the known model in (1) and (4); 
	\item Construct filter (2) with functions in (7) as fault estimator, given the known matrices in Step 3;
	\item Solve the semi-definite program in Theorem 1 to design the fault estimator parameters in Step 4.
\end{enumerate} 

\begin{rem}[\emph{\textbf{Limitations}}] \label{rem:limitation}
	{The limitations of the proposed method are listed as:}
	\begin{itemize}
		\item \textbf{Same Distribution Matrices for Fault and Disturbance}: If the system in (1) has the same fault and disturbance distribution matrices (i.e., the same $B_\omega$ and $B_f$ in (1a)), we cannot provide an accurate estimate of the fault (while exploiting robust control techniques, we cannot induce a low $\mathcal{L}_2$-gain with respect to same entry disturbance). This is a challenging open problem that is, in general, impossible to address without assuming known ``signal" characteristics of the fault and disturbance entering the dynamics through the same channel (e.g., frequency/power content, stochasticity, and even closed-form expressions for faults). We remark, however, that our approach allows for the estimation of combined fault and disturbance signals. Moreover, some existing work, such as \cite{van2022multiple}, explores this challenge for linear time-varying systems. In \cite{van2022multiple}, regression methods are used to isolate two same entry faults. It's important to note that this is different from fault estimation. The assumption made in \cite{van2022multiple} is that the faults are piece-wise constant and a similar condition of persistently excitation exists in one of the fault signals.
		\item \textbf{Pre-Defined Fault Estimation Performance}: By the proposed scheme, we cannot provide a pre-defined performance for fault estimation. We only minimize the effect of perturbations and find an optimal fault estimate in terms of the gains from disturbances and noise to fault estimation error. However, the accuracy of fault estimates might not be acceptable in some cases, depending on system characteristics and disturbance levels. 
	\end{itemize}
\end{rem}

In what follows, we discuss the uncertainty model and how different types of models affect the provided solution of the fault estimation problem. 
	
	\subsection{Discussion on Uncertainty Models}\label{sec:discussion_on_uncertainty_model}
	To obtain an uncertainty model of the form $\eta_{l_x}$ in \eqref{eq:eta_lx}, we can use results in e.g., \cite{quaghebeur2021incorporating, brunton2016discovering, yazdani2020systems}. Available methods allow fitting parametric static functions, which might be state or output dependent. To use these results in the context of fault estimation, we have to assume that there is some time window during the system operation in which no fault occurs and that data is collected for this healthy mode. The collected data can be used to learn uncertainty models to have a more accurate system description (valid at least for trajectories close to the training data set) for fault estimation.

	Based on \eqref{eq:eta_lx}, we have assumed that we have the state dependent uncertainty model $\eta_{l_x}(\cdot)$ to develop the results of this paper. In the case of an output dependent uncertainty model $\eta_{l_y}(\cdot)$, \eqref{eq:eta_lx} will change as follows:
	\begin{equation} \label{eq:eta_ly}
		\begin{aligned}
			\eta(V_\eta x, u,t) = \eta_{l_y}(T_\eta y, u,t) + \delta \eta_{l_y}(x, u,t), 
		\end{aligned}
	\end{equation}
where $\delta \eta_{l_y}(x, u,t) := \eta(V_\eta x, u,t) -  \eta_{l_y}(T_\eta y, u,t)$. Matrix $T_\eta$ is a selection matrix that the user selects to specify what particular outputs drive the model. If we use this $\eta_{l_y}(\cdot)$ instead of the state dependent $\eta_{l_x}(\cdot)$ in \eqref{eq:eta_lx}, \eqref{eq:augmented_matrices_fault} modifies as follows:  
		\begin{equation} \label{eq:augmented_matrices_fault_y}			
	\begin{aligned}
		B_{u_a} &:=	\left[\begin{array}{cc}
			{B}_u & S_\eta \\
			0 & 0\\
		\end{array}\right], \quad 
		u_a := \left[\begin{array}{c}
		u \\
		\eta_{l_y}(T_\eta y, u,t)
	\end{array}\right], \quad
		S_{g_a} := \left[\begin{array}{c}
			{S}_g\\
			0 
		\end{array}\right], \\
		V_{g_a} &:= {\left[\begin{array}{cc} V_g  & 0 \end{array}\right]},  \quad
		{g}_a(\cdot) :=  g(\cdot), \quad
		\omega_a :=	\left[\begin{array}{ccc}
			\delta \eta_{l_y}^T &
			\omega^T &
			f^{{(r)}^T}
		\end{array}\right]^T.
	\end{aligned}
\end{equation}
As you can see above (see $g_a(\cdot)$ in \eqref{eq:augmented_matrices_fault_y}), $\eta_{l_y}(\cdot)$ does not act as a nonlinearity for the augmented system (as $\eta_{l_x}$ does in \eqref{eq:augmented_matrices_fault}); instead, it acts as a known input signal. The latter affects the required assumption for the application of this uncertainty model (See Remark \ref{rem:uncertainty_models}).  

\begin{rem}[\emph{\textbf{Uncertainty Model Selection}}] \label{rem:uncertainty_models}
In Assumption \ref{assum:lipschitz}, the state-dependent uncertainty model $\eta_{l_x}$ must be globally Lipschitz to ensure the boundedness of the fault estimator. On the other hand, the output-dependent model $\eta_{l_y}$ relaxes this requirement by only requiring $\eta_{l_y}$ to be continuous. The latter comes at the price of lower model accuracy when the system uncertainty is not an explicit function of the system output (i.e, $V_\eta x$ might contain more states than those measured in the output).
\end{rem}
	\section{Simulation Results}   \label{sec:sim_results}
\begin{figure}[t!]
		\centering
		\smallskip
		\includegraphics[width=.6\linewidth,keepaspectratio]{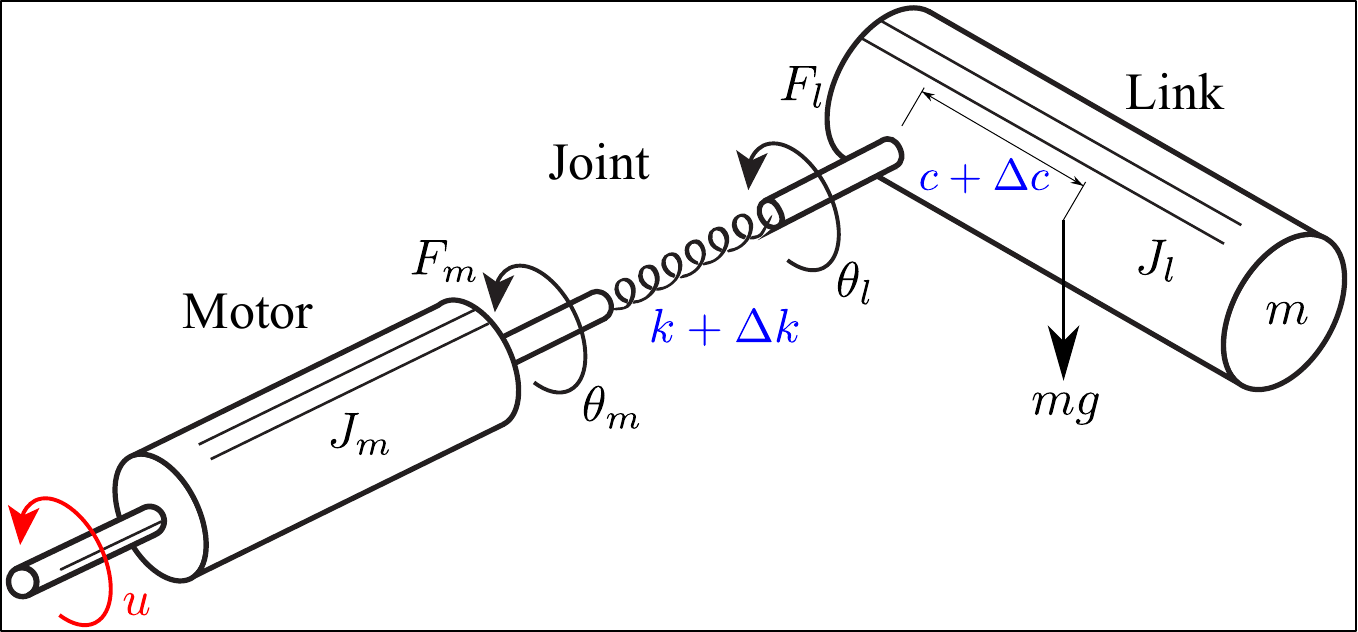}
		\caption{Benchmark System Schematic.}
		\label{fig:robot}
	\end{figure}

	In this section, we evaluate the proposed method using a benchmark example for FDI \cite{Keliris2017, reppa2013adaptive,zhang2005sensor, ghanipoor2022ultra}. The system dynamics (a single-link robotic arm with a revolute elastic joint, see Figure \ref{fig:robot} for a schematic) can be described as follows:
	\begin{equation*}
		\begin{aligned}
			& J_{b} \ddot{q}_{l}+F_{l} \dot{q}_{l}+(k_s+\Delta k_s)\left(q_{l}-q_{m}\right) +m g (c+ \Delta c) \sin \left(q_{l}\right) = \omega, \\
			& J_{m} \ddot{q}_{m}+F_{m} \dot{q}_{m}-(k_s+\Delta k_s)\left(q_{l}-q_{m}\right) =k_{\tau} u,
		\end{aligned}
	\end{equation*}
	where ${q}_{l}$ and ${q}_{m}$ are the angular position of the link and the angular position of the motor, respectively. Constants $J_{b}$ and $J_{m}$ are the moments of inertia of the link and the motor, and $F_{l}$ and $F_{m}$ are the viscous coefficients associated with friction acting at the link and the motor, respectively. The flexibility in the joint is modeled by a spring with a spring coefficient $k_s$, the inaccuracy in the spring coefficient is denoted by $\Delta k_s$, $m$ is the link mass, $g$ is the gravity constant, $c$ is the height of the link center of mass,  $\Delta c$ is the inaccuracy in the height of the link center of mass, $k_\tau$ is the amplifier gain, and $u$ is the torque input delivered by the motor. Units are in SI, and the parameters values are: $ J_{b} = 4.5$ $kgm^2, J_{m} = 1$ $kgm^2, F_{l} = 0.5$ $Nms/rad,F_{m} = 1$ $Nms/rad, k_s = 2$ $Nm/rad, \Delta k_s = -0.25k, m = 4$ $kg, g = 9.8$ $m/s^2, c = 0.5$ $m, \Delta c = 0.25c$ and $k_\tau = 1$. The torque input is set to $u = 2 \text{sin}(0.25t)$, and $\omega = 0.03 \text{sin}(0.1t)$ is an exogenous unknown (torque) disturbance affecting the link.
	
	Let $x_1 := \dot{q}_m, x_2 := q_m, x_3 := \dot{q}_l$, $x_4 := q_l $, and for the fault scenario, we consider the actuator fault; then, the system can be written in the form of \eqref{eq:sys}
	
	\begin{equation}
		\left\{\begin{aligned}
			\dot{x} &= A x+ B_u u+ S_g g\left(V_g x\right)+ S_\eta \eta \left(V_\eta x\right)  + B_\omega \omega + B_f f, \\
			y &= C x + D_f f+ D_\nu \nu,
			\label{eq:benchmark_sys}
		\end{aligned}\right.
	\end{equation}
	where $x := [x_1, x_2, x_3, x_4]^T$ is the state vector,
	\begin{equation*}
		\begin{aligned}
			A &=\left[\begin{array}{cccc}
				-\frac{F_{m}}{J m} & -\frac{k_s}{J_{m}} & 0 & \frac{k_s}{J_{m}} \\
				1 & 0 & 0 & 0 \\
				0 & \frac{k_s}{J_{l}} & -\frac{F_{l}}{J_{l}} & -\frac{k_s}{J_{l}} \\
				0 & 0 & 1 & 0
			\end{array}\right], \quad
			B_u=\left[\begin{array}{cccc}
			\frac{k_{\tau}}{J_{m}} & 0 & 0 & 0 \\
		\end{array}\right]^T, \quad
		C=\left[\begin{array}{cccc}
			0 & 1 & 0 & 0 \\
			0 & 0 & 0 & 1
		\end{array}\right], \\
		\end{aligned}
	\end{equation*}
	\begin{equation*}
		\begin{aligned}
			S_g &=\left[\begin{array}{cccc}
				0 & 0 & -\frac{m g c}{J_{l}} & 0 \\
			\end{array}\right]^T, \quad
			V_g=\left[\begin{array}{cccc}
				0 & 0 & 0 & 1 \\
			\end{array}\right], \quad
			S_\eta =\left[\begin{array}{cccc}
				1 & 0 & 0 & 0 \\
				0 & 0 & 1 & 0 \\
			\end{array}\right]^T,\\
			V_\eta&=\left[\begin{array}{cccc}
				0 & 1 & 0 & 0 \\
				0 & 0 & 0 & 1 \\
			\end{array}\right], \quad
			B_\omega=\left[\begin{array}{cccc}
				0 & 0 & 1 & 0 \\
			\end{array}\right]^T, \quad 
		B_f=\left[\begin{array}{llll}
			1 &
			0  &
			0  &
			0  
		\end{array}\right]^T,  \\
			D_f &=\left[\begin{array}{ll}
				0  &
				0 
			\end{array}\right]^T, \quad
		D_\nu= I. \\
		\end{aligned}
	\end{equation*}

	The nonlinearity is given by $g\left(Vx\right) =  \sin \left(x_{4}\right)$, which is Lipschitz with constant $\alpha = 1$. The uncertainty is $\eta(V_\eta x) =[
	\frac{\Delta k_s}{J_m} (x_4 - x_2),  \frac{\Delta k_s}{J_l} (x_2 - x_4) -\frac{m g \Delta c}{J_{l}} sin(x_4) 
	]^T $, which is induced by the uncertainty on the stiffness and the location of the center of mass of the link. We set initial conditions as $x(0) = [0.01,0.01,0.01,0.01]^T$. 
	
	We have the following linear state- and output-dependent uncertainty models:
	\begin{equation} \label{eq:eta_l_bench}
		\begin{aligned} 
			\eta_{l_x}(V_{\eta} {x}; \Theta_x)=&\Theta_{x} V_{\eta} {x}, \\
			\eta_{l_y}(T_\eta {y}; \Theta_y)=&{\Theta_{y}} T_\eta y,
		\end{aligned} 
	\end{equation}
	where $T_\eta$ is the identity matrix (since the uncertainties are dependent on both outputs). Matrices $\Theta_x$ and $\Theta_y$ are parameters of state- and output-dependent models with appropriate dimensions, respectively.

	Secondly, considering the system in \eqref{eq:benchmark_sys} and either of state and output dependent uncertainty models in \eqref{eq:eta_l_bench}, we construct augmented system \eqref{eq:augmented_system_fault} with $r=1$, using \eqref{eq:augmented_matrices_fault} or its modified version \eqref{eq:augmented_matrices_fault_y} for state and output dependent uncertainty models, respectively. Next, we design two actuator fault estimators of the form \eqref{eq:observer}, one for each of cases considering state- and output-dependent uncertainty models, by solving the semi-definite problem in Theorem \ref{theorem:fault_estimation}. The initial condition of the filters in simulation is taken as the zero vector. A sinusoidal actuator fault with the same frequency of input is simulated (i.e., $f_x = 0.1 sin(0.25(t-25))$). 
	
	We evaluate two aspects of the proposed methods: 
	\begin{enumerate}
		\item Effect of the learning models for uncertainty in a noise-free case. 
		\item Effect of robustification of fault estimate against noise.  
	\end{enumerate}
	\subsection{Model Learning for Uncertainty}
\begin{figure}[t!] 	
	 	\centering
	 	\smallskip
	 	\includegraphics[width=.7\linewidth,keepaspectratio]{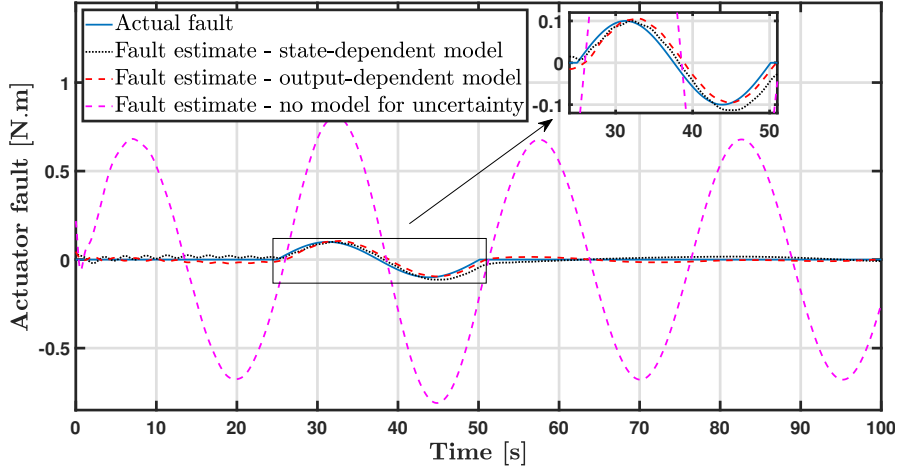}
		\caption{The actual actuator fault and its estimates.}
	 	\label{fig:estimated_fault}
	 \end{figure}

{ To indicate the performance of the proposed fault estimation approaches (two cases considering two uncertainty models) in a noise-free situation, we compare those approaches with the case in which we neither use the uncertainty model in the design of the fault estimator nor robustify the fault estimation error against perturbations induced by uncertainty (see the fault estimator given in \cite{ghanipoor2022ultra}). Figure \ref{fig:estimated_fault} depicts the actual fault and its estimates for three different approaches (two proposed methods plus one without any model for uncertainty). It can be seen that the estimated actuator faults using both proposed estimators follow the actual fault properly. Note that the fault estimate accuracy for this system does not change that much using either of the uncertainty models. However, this observation cannot be generalized since the fault estimation accuracy depends on the used uncertainty model (i.e., state- or output-dependent model) in the fault estimator filter and the gain from the uncertainty mismatch to fault estimate error. The combination of these two factors might not result in similar fault estimate accuracy for every system.
}

	\subsection{Robustification against Noise}
	
\begin{figure}[t!] 	
		\centering
		\smallskip
		\includegraphics[width=.7\linewidth,keepaspectratio]{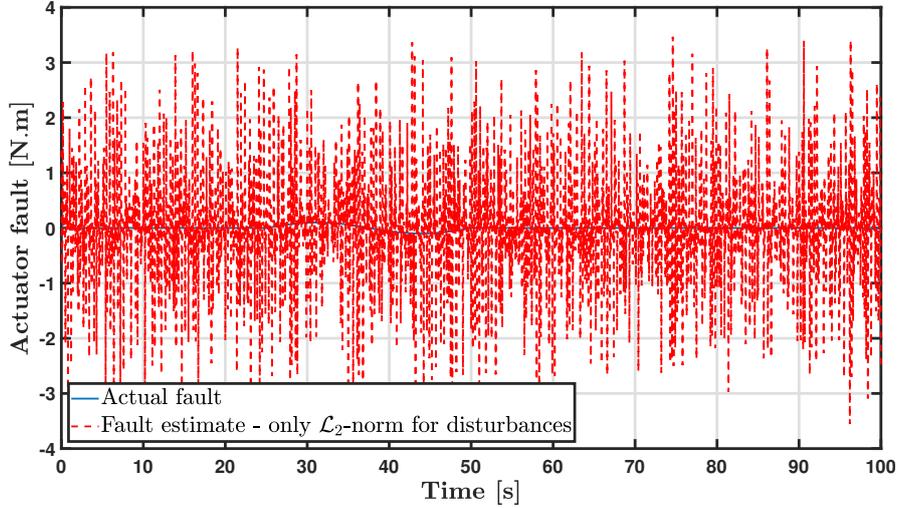}
		\caption{{The actual actuator fault and its estimates using the optimal $\mathcal{L}_2$-gain with respect to disturbance $\omega_a$ estimator.}}
		\label{fig:fault_estimate_noise_hinf}
	\end{figure}

\begin{figure}[t!] 	
		\centering
		\smallskip
		\includegraphics[width=.7\linewidth,keepaspectratio]{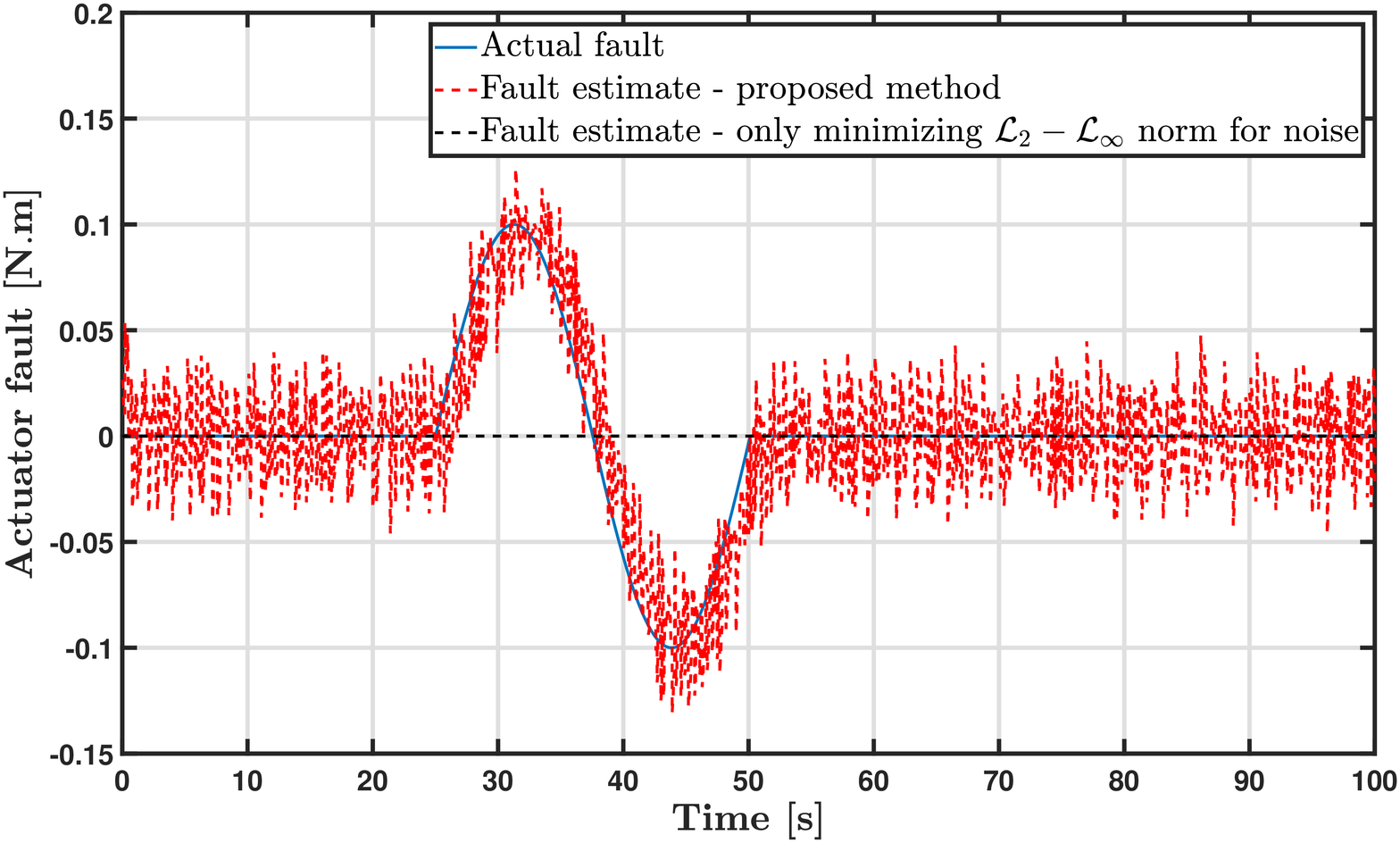}
		\caption{{The actual actuator fault and its estimates using different optimal criteria.}}
		\label{fig:fault_estimate_noise_h2}
	\end{figure}

	{In this section, we aim to show that the fault estimation performance and the noise attenuation are always a trade-off, and that the proposed synthesis approach allows us to make this trade-off in a constructive manner. Therefore, in the simulation of this section, the measurement noise $\nu$ is generated from a uniform distribution with an amplitude of five percent of the output signals for both sensors. Note that in the result of this section the modeling uncertainty is available and the state-dependent uncertainty model is used in the fault estimator. To show the trade-off between fault estimation performance and noise attenuation, we consider the filter of the form (7), which can be designed in three different ways as follows: }
	\begin{enumerate}
		\item Only minimizing the $\mathcal{L}_2$-gain from $\omega_a$ to $e_f$ in (10) (assuming $\nu_a = 0$), subject to the ISS LMI in (15a).
		\item {Only minimizing the $\mathcal{L}_2-\mathcal{L}_\infty$ induced gain from $\nu_a$ to $e_f$ in (10) (assuming $\omega_a = 0$), subject to the ISS LMI in (15a).}
		\item {Minimizing the $\mathcal{L}_2$-gain, subject to an upper bound for $\mathcal{L}_2-\mathcal{L}_\infty$ induced norm and the ISS LMI in (15a) (proposed method in Theorem 1).}
	\end{enumerate}
	{The first case of the above-mentioned scenarios is depicted in Figure \ref{fig:fault_estimate_noise_hinf}. One can see that the effect of noise is dominant in fault estimate, and this is due to the fact that by only minimizing $\mathcal{L}_2$-gain, we obtain a high-gain filter, which amplifies the noise effect. In contrast, by only minimizing $\mathcal{L}_2-\mathcal{L}_\infty$ induced norm, a low-gain filter is found, which (almost) perfectly filters noise effect but sacrifices fault estimation performance. Figure \ref{fig:fault_estimate_noise_h2} shows the result for this filter in dashed-black. The method proposed in this paper can provide a trade-off between the two previous solutions. Figure \ref{fig:fault_estimate_noise_h2} depicts the result using the proposed method in Theorem 1, see the dashed-red line. It can be observed that we have a decent trade-off between fault estimate performance and noise attenuation. Note that by tuning the upper bound for $\mathcal{L}_2-\mathcal{L}_\infty$ induced norm from $\nu_a$ to the fault estimate $e_f$, one can increase noise filtering at the cost of reducing fault estimate accuracy.}

	\section{Conclusion} \label{sec:conclusion}
	This paper proposes a method for the estimation of time-varying actuator and sensor faults in uncertain nonlinear systems. The fault estimator exploits an internal, ultra-local in time, model for the fault vector which allows us to guarantee zero fault estimation error for a class of faults in the absence of perturbations. The fault estimator can be designed by solving a semi-definite program. Herein, the effect of (fault and uncertainty) model mismatches, and external disturbances on the fault estimation error is minimized in the sense of $\mathcal{L}_2$-gain, for an acceptable $\mathcal{L}_2-\mathcal{L}_\infty$ induced norm with respect to measurement noise. This allows to design fault estimators that exhibit a favorable performance trade-off in the presence of these different perturbations challenging fault estimation. Simulations for a benchmark system illustrate the performance and potential of the proposed approach.

	\appendix
	\section*{Appendix}  
	\section{Proof of Proposition 1}  \label{ap:propos1_proof}  
	Let us first introduce the following lemma, which is used to ensure ISS using an ISS Lyapunov function.
	
	\begin{lem}\emph{\textbf{(ISS Lyapunov Function~{\cite[Thm. 4.19]{khalil2002nonlinear}})}
			Consider the error dynamics \eqref{eq:error_dynamics_combined} and let $W(e)$ be a continuously differentiable function such that
			\begin{equation*}
				\alpha_{1}(\|e\|) \leq W(e) \leq \alpha_{2}(\|e\|),
			\end{equation*}
			\begin{equation*}
				\dot{W}(e) \leq-W_{3}(e), \quad \hspace{-1mm} \forall \hspace{1mm}\|e\| \geq \xi (\| \bar \omega_a (t) \|),
			\end{equation*}
			where $\alpha_{1}(\cdot)$ and $\alpha_{2}(\cdot)$ are class $\mathcal{K}_{\infty}$ functions, $\xi(\cdot)$ is a class $\mathcal{K}$ function, and $W_{3}$ is a continuous positive definite function. Then, the estimation error dynamics \eqref{eq:error_dynamics_combined} is ISS with ISS gain $ \mu(\|\bar \omega_a\|) = \alpha_{1}^{-1}(\alpha_{2}(\xi(\|\bar \omega_a\|)))$.}
		\label{lem: iss}
	\end{lem}
	Let $W(e):={e}^{T} P {e}$ be an ISS Lyapunov function candidate. Then, it follows from \eqref{eq:error_dynamics_split} (equivalent to \eqref{eq:error_dynamics_combined}) and the Lipschitz conditions for the known nonlinearity in \eqref{eq:lipschitz} that
	\begin{equation}
		\begin{aligned} \dot{W}(e) \leq& e^{T} \Delta e  + \bar \omega_a^T \bar{T}^T M_\nu \bar{T} \bar \omega_a + 2 e^{T} P\bar B_{\omega_a} \bar \omega_a
		\end{aligned}
		\label{eq:lyapanov}
	\end{equation}
	with $\bar \omega_a, \bar{B}_{\omega_a}$ as defined in \eqref{eq:omega_a_bar}, 
	\begin{equation}
		\begin{aligned} 
			\Delta := &N^{T} P+P N+2\alpha P M S_{g_a} S_{g_a}^{T} M^{T} P +\alpha (V_{g_a} - J C_{a})^{T} (V_{g_a} - J C_{a}),
		\end{aligned}
		\label{eq:design_delta}
	\end{equation}
	and
	\begin{equation}
		\begin{aligned}
			M_\nu := T_\nu^T J^T  J T_\nu, \quad
			T_\nu := [\begin{array}[]{ll}  D_\nu & 0 \end{array}], \quad \bar{T} := [\begin{array}[]{ll} 0  & I_{2m} \end{array}].
		\end{aligned}
		\label{eq:Lyapanov_matrices}
	\end{equation}
	
	\begin{equation}		
		\begin{aligned} \dot{W} = &e^{T}\left(N^{T} P+P N\right) e+2 e^{T} P M S_{g_a} \delta g - 2 e^{T} P M B_{\omega_a} \omega_a + 2 e^{T} P B_{\nu_a} \nu_a\\
			\leq &e^{T}\left(N^{T} P+P N\right) e+2\left\|e^{T} P M S_{g_a} \right\|\|\delta g\| - 2 e^{T}  P M B_{\omega_a} \omega_a + 2 e^{T} P B_{\nu_a} \nu_a \\
			\leq &e^{T}\left(N^{T} P+P N\right) e +2\left\|e^{T} P M S_{g_a}\right\| \alpha\|(V_{g_a}  - J C_{a}) {e} + J D_\nu \nu \| - 2 e^{T} P M B_{\omega_a} \omega_a + 2 e^{T} P B_{\nu_a} \nu_a\\
			\leq &e^{T}\left(N^{T} P+P N\right) e +2\left\|e^{T} P M S_{g_a}\right\| \alpha (\|(V_{g_a}  - J C_{a}) {e} \| + \|J T_\nu \nu_a \|) - 2 e^{T} P M B_{\omega_a} \omega_a \\
			& + 2 e^{T} P B_{\nu_a} \nu_a\\
			\leq &e^{T}\left(N^{T} P+P N\right) e +\alpha\big(2\left\|e^{T} P M S_{g_a}\right\|^{2}+	\|(V_{g_a} - J C_{a}) e\|^{2} + \|J T_\nu \nu_a \|^2\big) - 2 e^{T} P M B_{\omega_a}  \omega_a \\
			&   + 2 e^{T} P B_{\nu_a} \nu_a  \\
			= &e^{T}\big(N^{T} P+P N+ 2\alpha P M S_{g_a} S_{g_a}^{T} M^{T} P +\alpha (V_{g_a} - J C_{a})^{T} (V_{g_a} - J C_{a}) \big) e 
		    + \nu_a^T \big( T_\nu^T J^T  J T_\nu \big) \nu_a \\
			&- 2 e^{T} P M B_{\omega_a} \omega_a + 2 e^{T} P B_{\nu_a} \nu_a\\ 
			= & e^{T} \Delta e  + \nu_a^T M_\nu \nu_a + 2 e^{T} P {\left[\begin{array}{cc}
					-M B_{\omega_{a}} & B_{\nu_a}
				\end{array}\right]} {\left[\begin{array}{cc}
					\omega_a	\\ \nu_a 
				\end{array}\right]} = e^{T} \Delta e  + \bar \omega_a^T \bar{T}^T M_\nu \bar{T} \bar \omega_a + 2 e^{T} P\bar B_{\omega_a} \bar \omega_a
		\end{aligned}
		\label{eq:lyapanov_ap}
	\end{equation}
	Now, inequality \eqref{eq:lyapanov} (which is the same as \eqref{eq:lyapanov_ap}) by taking the norm of the right-hand side of inequality implies the following inequality:
	\begin{equation}
		\begin{aligned} \dot{W}(e) \leq& -\lambda_{\min }(-\Delta) \|e\|^{2} + \|\bar{T}^T M_\nu \bar{T} \| \|\bar \omega_a\|^2 + 2 \|e\| \|P \bar  B_{\omega_a} \| \|\bar \omega_a\| \\
			=& - (1-\chi) \lambda_{\min }(-\Delta) \|e\|^{2} - \chi \lambda_{\min }(-\Delta) \|e\|^{2} + \|\bar{T}^T M_\nu \bar{T} \| \|\bar \omega_a\|^2  + 2 \|e\| \|P  \bar B_{\omega_a} \| \|\bar \omega_a\|\\
		\end{aligned}
		\label{eq:iss_analysis}
	\end{equation}
	for any $\chi \in (0,1)$ and $\lambda_{\min}(\cdot)$ the minimum eigenvalue of its symmetric argument. Now, a class $\mathcal{K}$ function $\xi(\bar \omega_a)$ exist such that we have  
	\begin{equation*}
		\begin{aligned} 
			\dot{W}(e) \leq& - (1-\chi) \lambda_{\min }(-\Delta) \|e\|^{2}, \quad \forall \hspace{1mm}\|e\| \geq \xi(\| \bar \omega_a \|),
		\end{aligned} 
	\end{equation*}
	$\xi(\bar \omega_a)$ exists since the second quadratic term in $\|e\|$ can dominate the third and fourth terms in \eqref{eq:iss_analysis} for large enough $\|e\|$. $\xi(\bar \omega_a)$ can be found by solving the following second-order inequality for $\|e\|$
	\begin{equation*}
		\begin{aligned}  \chi \lambda_{\min }(-\Delta) \|e\|^{2}  \geq &  \|\bar{T}^T M_\nu \bar{T} \| \|\bar \omega_a\|^2 + 2 \|e\| \|P \bar B_{\omega_a} \| \|\bar \omega_a\| 
		\end{aligned}
	\end{equation*}
	Since an explicit expression for the ISS gain is not needed in this paper, we do not give a closed-form solution for $\xi(.)$. 
	
	Therefore, the conditions in Lemma \ref{lem: iss} are satisfied if $\Delta$ is a negative definite matrix. Hence, under such condition system \eqref{eq:error_dynamics_combined} is ISS with input $\bar \omega_a$. Based on the above analysis, the proposition formalizes an LMI condition ($\Delta \prec 0$) that guarantees an ISS estimation error dynamics \eqref{eq:error_dynamics_combined}.
	
	As the final step, we want to prove that $\Delta \prec 0$ is equivalent to \eqref{eq:stability_lmi}-\eqref{eq:X12}. Using $\Delta$ defined in \eqref{eq:design_delta}, \eqref{eq:nuio_matrices_fault}, and the Schur complements on $\Delta \prec 0$, we can derive \eqref{eq:stability_lmi} with $X_{11}$ and $X_{12}$ in terms of the original observer gains $\{E,K,J\}$ as
	\begin{equation*}
		\begin{aligned}
			&X_{11} := N^{T} P+P N +\alpha (V_{g_a}^{T} V_{g_a} -V_{g_a}^{T}  J  C_a - C_a^{T}  J^{T}  V_{g_a}), \\
			&X_{12} :={\left[\begin{array}{ll}
					\sqrt{2\alpha} P M S_{g_a} &  \sqrt{\alpha} C_a^{T}  J^{T} 
				\end{array}\right]},
		\end{aligned}
	\end{equation*}
	where $N^{T} P+P N$ expands as
	\begin{equation*}
		\begin{aligned}
			&A_{a}^{T}\left(I+E C_{a}\right)^{T} P-C_{a}^{T}  K P+P\left(I+E C_{a}\right) A_{a}-P K C_{a},
		\end{aligned}
	\end{equation*}
	and we have
	\begin{equation*}
		\begin{aligned}
			P M S_{g_a} &= P S_{g_a}+P E C_{a} S_{g_a}.
		\end{aligned}
	\end{equation*}
	Consider the following change of variables
	\begin{equation*}
		R :=P E,  \qquad	Q :=P K.
		\label{eq:variable_change}
	\end{equation*}
	Applying this change of variables on the above expanded $X$ and $X_{12}$, the linear inequality \eqref{eq:stability_lmi}-\eqref{eq:X12} can be concluded. 
	
	When $\alpha$ is zero (no nonlinearity), it holds that $\Delta=N^T P + PN$ and, therefore, we only require $N^{T} P+P N$ to be negative definite, which by change of variables transforms to $S_{11} \prec 0$ is necessary and sufficient condition for ISS of the linear error dynamics \cite[Col. 5.1]{khalil2002nonlinear}. This conclude the results of the proposition.

	\section{Proof of Proposition \ref{propos:L_2_gain}} \label{ap: l_2_gain}
	Let us first introduce the following Lemma, which is required to prove the result of Proposition \ref{propos:L_2_gain}. In the following lemma, we state a Lyapunov-based sufficient condition for having such a bounded $\mathcal{L}_2$-gain property.
	\begin{lem} \emph{\textbf{($\mathcal{L}_2$-gain Inequality)}}
		Consider \eqref{eq:error_dynamics_split} with $\nu(t) = 0$ and suppose there exists a continuously differentiable positive semi-definite function $W(e)$ satisfying
		\begin{equation} \label{eq:hamilton-jacobi}
			\begin{aligned} 
				\dot{W}(e) &\leq a(\lambda^2 \omega_a^T \omega_a -  e_f^T e_f),
			\end{aligned}
		\end{equation}
		with $a, \lambda>0 $ and the fault estimation error $e_f$ as in \eqref{eq:error_def}. Then, the $\mathcal{L}_2$-gain from $\omega_a$ to $e_f$ in \eqref{eq:error_dynamics_split} is less than or equal to $\lambda$.
		\label{lem:l2_gain}
	\end{lem}
	\emph{	\textbf{Proof}: } The proof is similar to the proof of Theorem 5.5 in \cite{khalil2002nonlinear}. Clearly, by integrating \eqref{eq:hamilton-jacobi} over a finite time $(0,\tau)$, we have 
	\begin{equation*}
		W(e(\tau))-W(e(0)) \leq a \lambda^2 \int_{0}^{\tau} \|\omega_a\|^2 dt - a \int_{0}^{\tau} \|e_f\|^2 dt.
	\end{equation*}
	Using $W(e) \succeq 0$ and neglecting the initial condition, the above inequality can be written as 
	\begin{equation*}
		\int_{0}^{\tau} \|e_f\|^2 dt \leq \lambda^2 \int_{0}^{\tau} \|\omega_a\|^2 dt 
	\end{equation*}
	Considering Definition \ref{def:l2-gain}, the above inequality concludes the result in the lemma. 
	\hfill $\blacksquare$
	
	For the purpose of $\mathcal{L}_2$-gain analysis, we define $W(e) :={e}^{T} P {e}$ with positive definite matrix $P$. Then, we can guarantee that the $\mathcal{L}_2$-gain inequality in \eqref{eq:hamilton-jacobi} holds for the time-derivative $\dot{W}(e)$ evaluated along solutions of the error-dynamics \eqref{eq:error_dynamics_split}, by using \eqref{eq:lyapanov_ap} (and assuming $\nu = 0$), as follows:
	\begin{equation*}
		\begin{aligned} 
			\dot{W}(e) &\leq e^{T} \Delta e  - 2 e^{T} P M B_{\omega_a} \omega_a \leq a(\lambda^2 \omega_a^T \omega_a  -   e^{T} \bar{C}^{T} \bar{C} e).
		\end{aligned}
	\end{equation*}
	The above inequality can be written as 
	\begin{equation*}
		e^{T}(\Delta + a \bar{C} ^{T} \bar{C} ) e  - 2 e^{T} P M B_{\omega_a} \omega_a - a \lambda^2 \omega_a^T \omega_a \leq 0,
	\end{equation*}
	for which we can give the following sufficient matrix inequality 
	\begin{equation*}
		\left[\begin{array}{cc}
			\Delta + a \bar{C} ^{T} \bar{C} & -P M B_{\omega_a} \\
			* & -a \lambda^2 I
		\end{array}\right]\preceq 0.
	\end{equation*}
	If we follow the same procedure in the proof of Proposition \ref{propos:stability}, the equivalent inequality can be found as
	\begin{equation*}
		\left[\begin{array}{ccc}
			X+ a \bar{C} ^{T} \bar{C} &  -(P + R C_a) B_{\omega_a}  & X_{12} \\
			* & -a \lambda^2 I  & 0 \\
			* & * & -I
		\end{array}\right]\preceq0.
	\end{equation*}
	Finally, by defining a change of variable as $\rho := \lambda^2$, the LMI condition \eqref{eq:l2_gain_lmi} implies the above inequality.
	
	When $\alpha$ is zero (linear case), the LMI condition \eqref{eq:l2_gain_lmi} using the Schur complement is equivalent to \eqref{eq:l2_gain_lmi_linear}. This LMI is necessary and sufficient condition for the the $\mathcal{L}_2$-gain of \eqref{eq:error_dynamics_split} (with $\delta g = 0$ for this linear case) from $\omega_a$ to the fault estimation error $e_f$ to be upper bounded by $\sqrt{\rho}$, as given in \cite[Prop. 3.12]{scherer2000linear}. 
	
	\section{Proof of Proposition \ref{propos:L2_Linf_norm}} \label{ap:l2_linf_norm}
	For the purpose of $\mathcal{L}_2-\mathcal{L}_\infty$-gain analysis, we define $W(e) :={e}^{T} P {e}$ with positive definite matrix $P$. Then, we can impose the quadratic performance inequality 
	\begin{equation}
		\begin{aligned} 
			\dot{W}(e) &\leq b^2 \nu_a^T \nu_a.
		\end{aligned}
		\label{eq:wdot_l2_linf}
	\end{equation}
	Using \eqref{eq:lyapanov_ap} (and assuming $\omega_a = 0$), \eqref{eq:wdot_l2_linf} can be written as follows:
	\begin{equation*}
		\begin{aligned} 
			\dot{W}(e) &\leq e^{T} \Delta e  + 2 e^{T} P B_{\nu_a} \nu_a + \nu_a^T M_\nu \nu_a \leq b^2 \nu_a^T \nu_a.
		\end{aligned}
	\end{equation*}
	The above inequality can be written as 
	\begin{equation*}
		e^{T}\Delta e  + 2 e^{T} P B_{\nu_a} \nu_a  + (M_\nu-b^2I) \nu_a^T \nu_a \leq 0,
	\end{equation*}
	for which we can give the following sufficient matrix inequality 
	\begin{equation*}
		\left[\begin{array}{cc}
			\Delta  & P B_{\nu_a} \\
			* & M_\nu-b^2 I
		\end{array}\right]\preceq 0.
	\end{equation*}
	Since $M_\nu = T_\nu^T J^T  J T_\nu$ is not linear in the design variable $J$, we use the Schur complement again to obtain the following equivalent inequality:
	\begin{equation*}
		\left[\begin{array}{ccc}
			\Delta &  P B_{\nu_a} & 0 \\
			* & -b^2 I  &   T_\nu^T J^T \\
			* & * & -I
		\end{array}\right]\preceq0.
	\end{equation*}
	If we follow the same procedure in the proof of Proposition \ref{propos:stability}, the equivalent LMI can be found:
	\begin{equation*}
		\left[\begin{array}{cccc}
			X & H_{12} & 0 & X_{12}\\
			* & -b^2 I  &  T_\nu^T J^T & 0\\
			* & * & -I & 0\\
			* & * & * & -I
		\end{array}\right]\preceq0,
	\end{equation*}
	where $H_{12}$ as defined in \eqref{eq:h12}. This completes the proof that the condition \eqref{eq:l2_linf_1} in the proposition imply the satisfaction of \eqref{eq:wdot_l2_linf}. Besides, by integrating \eqref{eq:wdot_l2_linf} over a finite time $(0,\tau)$ and neglecting the initial condition, we have 
	\begin{equation*}
		{e}(\tau)^{T} P {e}(\tau) \leq b^2 \int_{0}^{\tau}  \nu_a^T \nu_a dt.
	\end{equation*}
	If we multiply both sides by $\gamma^2$ and impose a lower bound to enable bounding the error on the fault estimate ($e_f = \bar{C} e$), we have 
	\begin{equation}
		{e}^{T}(\tau) {\bar{C}}^{T} \bar{C} e(\tau)	\leq \gamma^2 {e}^{T}(\tau) P {e}(\tau) \leq  (\gamma b)^2 \int_{0}^{\tau}  \nu_a^T \nu_a dt.
		\label{eqapp:l2_linf_ineq}
	\end{equation}
	Then, it follows from the first part of the above inequality that we need
	\begin{equation*}
		{e}^{T}(\tau) (P- \frac{{\bar{C}}^{T} \bar{C}}{\gamma^2}) e(\tau)	\geq 0.
	\end{equation*}
	Using the Schur complement, the following LMI implies the satisfaction of the above inequality:
	\begin{equation*}
		\left[\begin{array}{cc}
			P	 & \bar{C}^T \\
			* & \sigma I
		\end{array}\right] \succeq 0,
	\end{equation*}
	where $\sigma := \gamma^2$.
	Now, by imposing the above LMI, \eqref{eqapp:l2_linf_ineq} indeed holds. Therefore, 
	\begin{equation*}
		\begin{aligned}
			\sup_{\tau\geq 0} {e}^{T}(\tau) {\bar{C}}^{T} \bar{C} e(\tau) &\leq  \sup_{\tau\geq 0} (\gamma b)^2 \int_0^\infty \nu_a^T \nu_a dt = (\gamma b)^2 \int_0^\infty \nu_a^T \nu_a dt \\
		\end{aligned}
	\end{equation*}
	which concludes the result in the proposition for nonlinear case.
	
	When $\alpha$ is zero (linear case), the LMI condition \eqref{eq:l2_linf_1} using the Schur complement is equivalent to \eqref{eq:l2_linf_gain_lmi_linear}. Then, \eqref{eq:l2_linf_gain_lmi_linear} and \eqref{eq:l2_linf_2} are necessary and sufficient conditions for the $\mathcal{L}_2-\mathcal{L}_\infty$ induced norm of \eqref{eq:error_dynamics_split} (with $\delta g = 0$ for this linear case) from $\nu_a$ to the fault estimation error $e_f$ to be upper bounded by $\sqrt{\sigma}$, as given in \cite[Prop. 3.15]{scherer2000linear}.  
	
	\section*{Acknowledgements}                               
	This publication is part of the project Digital Twin project
	4.3 with project number P18-03 of the research programme
	Perspectief which is (mainly) financed by the Dutch Research
	Council (NWO). P. Mohajerin Esfahani also acknowledges the support of the European Research Council (ERC) under the European Unions Horizon 2020 research and innovation programme (TRUST-949796).
	
	\bibliographystyle{unsrt}        
	\bibliography{ref}           
	
\end{document}